\documentclass[onecolumn,amsmath,showpacs,nofootinbib,12pt]{revtex4-2}
\usepackage{graphicx}
\usepackage{multirow}
\usepackage{dcolumn}
\usepackage{bm}
\usepackage{color}
\begin{document}
\newcommand{\hs}{\hspace*{0.5cm}}
\newcommand{\vs}{\vspace*{0.5cm}}
\newcommand{\be}{\begin{equation}}
\newcommand{\ee}{\end{equation}}
\newcommand{\bea}{\begin{eqnarray}}
\newcommand{\eea}{\end{eqnarray}}
\newcommand{\ben}{\begin{enumerate}}
\newcommand{\een}{\end{enumerate}}
\newcommand{\bde}{\begin{widetext}}
\newcommand{\ede}{\end{widetext}}
\newcommand{\nn}{\nonumber}
\newcommand{\crn}{\nonumber \\}
\newcommand{\Tr}{\mathrm{Tr}}
\newcommand{\non}{\nonumber}
\newcommand{\noi}{\noindent}
\newcommand{\al}{\alpha}
\newcommand{\la}{\lambda}
\newcommand{\bet}{\beta}
\newcommand{\ga}{\gamma}
\newcommand{\va}{\varphi}
\newcommand{\om}{\omega}
\newcommand{\pa}{\partial}
\newcommand{\+}{\dagger}
\newcommand{\fr}{\frac}
\newcommand{\bc}{\begin{center}}
\newcommand{\ec}{\end{center}}
\newcommand{\Ga}{\Gamma}
\newcommand{\de}{\delta}
\newcommand{\De}{\Delta}
\newcommand{\ep}{\epsilon}
\newcommand{\varep}{\varepsilon}
\newcommand{\ka}{\kappa}
\newcommand{\La}{\Lambda}
\newcommand{\si}{\sigma}
\newcommand{\Si}{\Sigma}
\newcommand{\ta}{\tau}
\newcommand{\up}{\upsilon}
\newcommand{\Up}{\Upsilon}
\newcommand{\ze}{\zeta}
\newcommand{\ps}{\psi}
\newcommand{\Ps}{\Psi}
\newcommand{\ph}{\phi}
\newcommand{\vph}{\varphi}
\newcommand{\Ph}{\Phi}
\newcommand{\Om}{\Omega}
\newcommand{\AdrHEPC}{Phenikaa Institute for Advanced Study and Faculty of Basic Science, Phenikaa University, Yen Nghia, Ha Dong, Hanoi 100000, Vietnam}

\title{Novel effects of the $W$-boson mass shift in the 3-3-1 model}

\author{Duong Van Loi}
\email{Corresponding author; loi.duongvan@phenikaa-uni.edu.vn}
\author{Phung Van Dong}
\email{dong.phungvan@phenikaa-uni.edu.vn}
\affiliation{\AdrHEPC} 

\date{\today}

\begin{abstract}
The recent precision measurement of the $W$-boson mass reveals an exciting hint for the new physics as of the 3-3-1 model. We indicate that the 3-3-1 model contains distinct sources by itself that may cause the $W$-mass deviation, as measured, such as the tree-level $Z$-$Z'$ mixing, the tree-level $W$-$Y$ and $Z$-$Z'$-$X$ mixings, as well as the non-degenerate gauge vector $(X,Y)$ and new Higgs doublets. We point out that the gauge vector doublet negligibly contributes to this mass shift, whereas the rest of the effects with tree-level mixings governed by $Z$-$Z'$ and new Higgs doublets are significant. A discussion of scalar sextet contributions is also given.  
\end{abstract}

\maketitle

\section{\label{intro}Introduction}
The CDF collaboration has recently announced a new result of $W$-boson mass $m_W|_{\mathrm{CDF}}=80.4335\pm 0.0094$ GeV \cite{CDF:2022hxs} which deviates from the standard model prediction $m_W|_\mathrm{SM}=80.357\pm 0.006$ GeV \cite{ParticleDataGroup:2020ssz} at $7\sigma$. Such a high precision measurement of $W$ mass may be a significant indication for the new physics beyond the standard model. 

On theoretical grounds, the CDF $W$-mass anomaly possibly originates from (i) a non-minimal Higgs sector that contains the standard model Higgs field and directly contributes to this mass deviation via relevant Higgs mechanism, (ii) tree level mixings of the standard model $Z$ and even $W$ bosons with new particles that cause the $W$ mass as shifted, and/or (iii) loop-level quantum corrections due to the presence of new particles to gauge boson self-energies that modify the Peskin--Takeuchi parameters $S,T,U$ \cite{Peskin:1990zt,Peskin:1991sw,Maksymyk:1993zm}. Several efforts have been made in the literature to provide possible solutions to this puzzle, see Refs. \cite{Lu:2022bgw,Du:2022pbp,deBlas:2022hdk,Athron:2022qpo,Strumia:2022qkt,Cacciapaglia:2022xih,Liu:2022jdq,Bagnaschi:2022whn,Asadi:2022xiy,Athron:2022isz,DiLuzio:2022xns,Song:2022xts,Gu:2022htv,Babu:2022pdn,Paul:2022dds,Biekotter:2022abc,Balkin:2022glu,Cheung:2022zsb,Du:2022brr,Heo:2022dey,Ahn:2022xeq,Kawamura:2022uft,Kanemura:2022ahw,Nagao:2022oin,Mondal:2022xdy,Carpenter:2022oyg,Zhang:2022nnh,Popov:2022ldh,Arcadi:2022dmt,Chowdhury:2022moc,Borah:2022obi,Zeng:2022lkk,Ghorbani:2022vtv,Baek:2022agi,Cao:2022mif,Borah:2022zim,Almeida:2022lcs,Cheng:2022aau,Addazi:2022fbj,Heeck:2022fvl,Cai:2022cti,Batra:2022pej,Benbrik:2022dja,Gupta:2022lrt,Lee:2022gyf,Wang:2022dte,VanDong:2022rox,Rodriguez:2022wix} for an incomplete list. In this work, we show that the model based upon $SU(3)_C\otimes  SU(3)_L\otimes U(1)_X$ (called 3-3-1) gauge symmetry \cite{Pisano:1992bxx,Frampton:1992wt,Foot:1992rh,Valle:1983dk,Montero:1992jk,Foot:1994ym} manifestly accommodates the CDF $W$-mass anomaly. The reason for this model choice is that all the effects dedicated above of the new physics are actually dictated by the 3-3-1 gauge principle, and thus they are very predictive. 

One of the fundamental motivations for the 3-3-1 gauge symmetry extension is that it can address the number of fermion families naturally. Indeed, in the standard model, the number of fermion families on the theoretical ground is left arbitrarily. The reason may come from a fact that the gauge anomalies are cancelled out for every fermion family; no correlation between families is needed. This is due to the weak isospin symmetry with the relevant $SU(2)_L$ trace $\mathrm{Tr}[\{T_i,T_j\}T_k]=0$ for any fermion representation. The simplest extension of $SU(2)_L$ to $SU(3)_L$ yields the corresponding trace nontrivially, which does not vanish for complex representations. With enlargement of respective fermion representations under $SU(3)_L$, each family now depends on anomaly. The $[SU(3)_L]^3$ anomaly vanishes if all the families must be taken into account, implying that the number of families is a multiple of color number (cf. \cite{Frampton:1992wt}). The QCD asymptotic freedom demands that the number of families is smaller than or equal to five. It follows that the family number is just three, coinciding with experiment. Hence, the core of the 3-3-1 extension is to signify a higher weak isospin symmetry, $SU(3)_L$, directly enlarged from $SU(2)_L$. It is noted that the new $U(1)_X$ group necessarily included ensures an algebraic closure between electric charge and $SU(3)_L$, analogously to the standard model that requires the hypercharge. The remaining anomalies associated with $U(1)_X$ vanish too. Interestingly, besides the fermion family number, this new gauge principle yields a potential solution to the profound questions of electric charge quantization \cite{Pisano:1996ht,Doff:1998we,deSousaPires:1998jc,deSousaPires:1999ca,Dong:2005ebq}, discrepancy of third quark family \cite{Ng:1992st,GomezDumm:1993oxo,Long:1999ij}, and strong CP conservation \cite{Pal:1994ba,Dias:2003iq,Dias:2003zt,Dong:2012bf}. Additionally, it naturally addresses the issues of neutrino mass generation \cite{Tully:2000kk,Dias:2005yh,Chang:2006aa,Dong:2006mt,Dong:2008sw,Dong:2010gk,Dong:2010zu,Dong:2011vb,Boucenna:2014ela,Boucenna:2014dia,Boucenna:2015zwa,Okada:2015bxa,Pires:2014xsa,Dias:2010vt,Huong:2016kpa,Reig:2016tuk}, flavor physics \cite{GomezDumm:1994tz,Buras:2012dp,Buras:2013dea,Gauld:2013qja,Buras:2014yna,Buras:2015kwd,Dong:2015dxw}, dark matter stability \cite{Fregolente:2002nx,Hoang:2003vj,Filippi:2005mt,deS.Pires:2007gi,Mizukoshi:2010ky,Alvares:2012qv,Profumo:2013sca,Kelso:2013nwa,daSilva:2014qba,Dong:2013ioa,Dong:2014esa,Dong:2015rka,Dong:2013wca,Dong:2014wsa,Dong:2015yra,Huong:2016ybt,Alves:2016fqe,Ferreira:2015wja,Dong:2017zxo}, as well as cosmic inflation and baryon asymmetry \cite{Huong:2015dwa,Dong:2018aak,Dong:2017ayu}. 

The 3-3-1 model can be classified, based upon the embedding of electric charge operator in the new gauge symmetry, say $Q=T_3+\beta T_8+X$, through the $\beta$ parameter, where $T_j$ ($j=1,2,3,\cdots,8$) and $X$ are $SU(3)_L$ and $U(1)_X$ charges, respectively. Generally, the 3-3-1 model possesses a lepton triplet of form $(\nu_L, e_L, E^q_L)$ where $E$ is some field with electric charge $q$, related to the $\beta$ parameter as $\beta=-(1+2q)/\sqrt{3}$. Notice that switching representations with conjugated representations of $SU(3)_L$, e.g. $(\nu_L,e_L, E^q_L)\to (e_L,-\nu_L,E^q_L)$, changes $\beta\to -\beta$ and leads to a version with rather similar phenomenology, including the $W$ mass, thus not interpreted in this work. There are two typical variants of the 3-3-1 model as far as the lepton sectors are relevant. The minimal 3-3-1 model combines known leptons into a triplet $(\nu_L,e_L,e^c_R)$ for each family, thus $\beta=-\sqrt{3}$ \cite{Pisano:1992bxx,Frampton:1992wt,Foot:1992rh}, whereas the 3-3-1 model with right-hand neutrinos introduces three right-handed neutrinos ($\nu_R$'s) to perform $(\nu_L,e_L,\nu^c_R)$ for each family, thus $\beta=-1/\sqrt{3}$ \cite{Valle:1983dk,Montero:1992jk,Foot:1994ym}. Particularly-interested theories that modify the minimal 3-3-1 model but keep $\beta=-\sqrt{3}$ consist of the 3-3-1 model with exotic charged leptons \cite{Pleitez:1992xh}, the reduced 3-3-1 model \cite{Ferreira:2011hm}, and the simple 3-3-1 model \cite{Dong:2014esa}. Also, the theories that modify the 3-3-1 model with right-handed neutrinos but keep $\beta=-1/\sqrt{3}$ include the economical 3-3-1 model \cite{Ponce:2002sg,Dong:2006mg} and the 3-3-1 model with neutral (heavy) fermions~\cite{Dong:2010gk,Dong:2010zu,Dong:2011vb,Mizukoshi:2010ky}.\footnote{Here the third entry of a lepton triplet is a new neutral fermion different from the usual right-handed neutrino as proposed long ago \cite{Singer:1980sw}, and especially flavor symmetries that determine lepton mixing naturally work in this kind of the 3-3-1 model.} Alternatively, the 3-3-1 model without exotic charge proposes new copies of charged leptons, called heavy charged leptons, forming $(\nu_L, e_L, E^-_L)$ for each family, thus $\beta=1/\sqrt{3}$ \cite{Pleitez:1994pu,Ozer:1995xi}. And, the flipped 3-3-1 model presents a distinct arrangement for fermion representations with $\beta=1/\sqrt{3}$ \cite{Fonseca:2016tbn,Huong:2019vej}. From this point of view, the variants of the 3-3-1 model mainly differ in fermion content and Higgs sector, besides $\beta$ that specifies the gauge spectrum. Although the fermion content is fixed by the anomaly cancelation, the QCD asymptotic freedom, and the electric charge embedding, the Higgs sector is actually arbitrary, having plenty of multiplets by contrast, as seen from the modified models of the typical 3-3-1 models. According to the existing 3-3-1 theories, each of which may possess two Higgs triplets, three Higgs triplets, three Higgs triplets plus one Higgs sextet, or even many Higgs multiplets if given a flavor symmetry. 

It is noted that the same electric charge embedding, i.e. $\beta$ (or $q$), cannot distinguish alternative particle contents and vacuum structures, e.g. the 3-3-1 model with right-handed neutrinos vs. the 3-3-1 model with neutral fermions, as well as unwanted Higgs vacuum alignments in these models. The behavior of baryon number minus lepton number ($B-L$) symmetry put forward in \cite{Dong:2013wca,Dong:2013ioa,Dong:2014esa,Dong:2015yra} for this kind of the model may provide insight in signifying the alternative particle contents as well as classifying the nontrivial Higgs vacuum structures. Hence, the 3-3-1 model can be characterized by $\beta$ (or $q$) and $B-L$ behavior, which are important for analysing the sources of the $W$-mass shift. As a matter of fact, the 3-3-1 models possibly contain new neutral ($Z',X$) and charged ($Y$) gauge bosons presenting interesting mixing phenomena with usual $W,Z$ bosons, which along with the Higgs vacuum structures cause the $W$-mass shift at tree level. Additionally, the new non-Hermitian gauge vector doublet $(X,Y)$ and inert Higgs multiplets presented in 3-3-1 models for neutrino mass and/or dark matter also contribute to this mass shift at loop level.     

The rest of this work is organized as follows. In Sec. \ref{model} we set up a generic 3-3-1 model in which relevant Higgs mechanism important for the $W$-mass shift is determined and classified by electric charge conservation and $B-L$ behavior.  In Sec. \ref{Wmass}, we investigate various novel contributions of the model to the $W$-mass shift. A remark of scalar sextet contribution to the $W$-mass shift is given in Sec. \ref{sextet}. The extra important constraints for 3-3-1 model are discussed in Sec. \ref{Constrain}. We make a conclusion in Sec. \ref{conclusion}.

\section{\label{model} Description of the model}

We first present the necessary features of the 3-3-1 model with arbitrary $\beta$ embedding (or $q$ charge) parameter. We then determine distinct gauge-boson mass spectra according to profiles of Higgs vacuum structures, which affect differently to the $W$-mass anomaly.

\subsection{Particle content}

The 3-3-1 gauge symmetry is given by \be SU(3)_C\otimes SU(3)_L\otimes U(1)_X,\label{gt1} \ee where the first factor is the usual color group, while the last two are directly extended from the electroweak group, as mentioned. The decomposition scheme of the extended gauge sector into the usual gauge groups takes the form, \be SU(3)_L\otimes U(1)_X\rightarrow SU(2)_L\otimes U(1)_{T_8}\otimes U(1)_X\rightarrow SU(2)_L\otimes U(1)_Y\rightarrow U(1)_Q,\ee where $T_j\,(j=1,2,3,\cdots,8)$ and $X$ stand for $SU(3)_L$ and $U(1)_X$ charges, respectively. Additionally, the hypercharge $Y$ and the electric charge $Q$ are embedded, respectively, as \bea Y &\equiv& \beta T_8+X,\label{gt3}\\  
Q &\equiv& T_3+Y=T_3+\beta T_8+X.\label{gt2}\eea  In other words, when the 3-3-1 symmetry is broken down to the standard model symmetry, the hypercharge is composed of the two new diagonal charges, $T_8$ and $X$, as broken. When the electroweak symmetry is broken down to electromagnetic symmetry, the electric charge is composed of the third weak-isospin component and the hypercharge, as usual. The coefficient $\beta$ is related to a basic charge parameter $q$ that is the electric charge of the third component ($E$) of a lepton triplet, such as $\beta = -(1+2q)/\sqrt{3}$. It is noted that the 3-3-1 model can possess variants that differ by corresponding $\beta$ (or $q$) values, as imposed in Table \ref{331versions}.  

\begin{table}[h]
\begin{tabular}{c|cccc}
\hline\hline
3-3-1 model & Minimal & RHNs & HCLs & Generic \\
\hline 
$q$ & $1$ & $0$ & $-1$ & $q\neq 1,0,-1$ \\
$\beta$ & $-\sqrt{3}$ & $-1/\sqrt{3}$ & $1/\sqrt{3}$ & $\beta\neq -\sqrt{3},\mp1/\sqrt{3}$ \\
Lepton content & $\begin{pmatrix}\nu \\ e \\ e^c \end{pmatrix}$ vs. $\begin{pmatrix}\nu \\ e \\ P \end{pmatrix}$ & $\begin{pmatrix}\nu \\ e \\ \nu^c \end{pmatrix}$ vs. $\begin{pmatrix}\nu \\ e \\ N \end{pmatrix}$ & $\begin{pmatrix}\nu \\ e \\ E^-\end{pmatrix}$ vs. Flip. & $\begin{pmatrix}\nu \\ e\\ E^q\end{pmatrix}$ \\
Scalar structure & $\left[\begin{array}{ccc}
\rho^+_1 & \eta^0_1 & \chi^-_1 \\
\rho^0_2  & \eta^-_2  & \chi^{--}_2 \\
\rho^{++}_3 & \eta^{+}_3 & \chi^{0}_3 \end{array}\right]$ & $\left[\begin{array}{ccc}
\rho^+_1 & \eta^0_1 & \chi^0_1 \\
\rho^0_2  & \eta^-_2  & \chi^{-}_2 \\
\rho^{+}_3 & \eta^{0}_3 & \chi^{0}_3 \end{array}\right]$ & $\left[\begin{array}{ccc}
\rho^+_1 & \eta^0_1 & \chi^+_1 \\
\rho^0_2  & \eta^-_2  & \chi^{0}_2 \\
\rho^{0}_3 & \eta^{-}_3 & \chi^{0}_3 \end{array}\right]$ & $\left[\begin{array}{ccc}
\rho^+_1 & \eta^0_1 & \chi^{-q}_1 \\
\rho^0_2  & \eta^-_2  & \chi^{-q-1}_2 \\
\rho^{q+1}_3 & \eta^{q}_3 & \chi^{0}_3 \end{array}\right]$ \\ 
Standard vacuum & $\left[\begin{array}{ccc}
0 & * & 0 \\
* & 0 & 0 \\
0 & 0 & * \end{array}\right]$ & $\left[\begin{array}{ccc}
0 & * & 0 \\
* & 0 & 0 \\
0 & 0 & * \end{array}\right]$ & $\left[\begin{array}{ccc}
0 & * & 0 \\
* & 0 & 0 \\
0 & 0 & * \end{array}\right]$ & $\left[\begin{array}{ccc}
0 & * & 0 \\
* & 0 & 0 \\
0 & 0 & * \end{array}\right]$ \\
Abnormal vacuum & No & $\left[\begin{array}{ccc}
0 & * & * \\
* & 0 & 0 \\
0 & * & * \end{array}\right]$ & $\left[\begin{array}{ccc}
0 & * & 0 \\
* & 0 & * \\
* & 0 & * \end{array}\right]$ & No\\
Landau pole & 4--5 TeV & $>M_{\mathrm{Pl}}$ & $>M_{\mathrm{Pl}}$ & Exist for large $|\beta|$\\
\hline\hline 
\end{tabular}
\caption[]{\label{331versions} A roadmap for 3-3-1 versions and relevant vacuum structures. Minimal: The minimal 3-3-1 model and its variants for $\beta=-\sqrt{3}$ (or $q=1$); RHNs: The 3-3-1 model with right-handed neutrinos and its variants for $\beta=-1/\sqrt{3}$ (or $q=0$); HCLs: The 3-3-1 model with heavy charged leptons and its flipped 3-3-1 variant (see \cite{Fonseca:2016tbn,Huong:2019vej} in detail) for $\beta=1/\sqrt{3}$ (or $q=-1$); Generic: The generic 3-3-1 model whose $\beta$ (or $q$) satisfies $|\beta|<1.824$ (or $-2.08<q<1.08$), except for the previous cases. [Verify Sec. \ref{Constrain} for the $\beta$ range and Landau pole.] A star ``$*$'' shows a vacuum expectation value viably for the corresponding scalar component. For $q=0,-1$, the standard vacuum applies for $N,E^-$ versions with $[B-L](N,E^-)=0$, protected by $P_M$, whereas the abnormal vacuum happens for $\nu^c,E^-$ versions with $[B-L](\nu^c,E^-)=1$, not protected by $P_M$.}
\end{table}

The fermion content transforms under the gauge symmetry in (\ref{gt1}) as
\bea && \psi_{aL} = 
\begin{pmatrix}
\nu_{aL}\\
e_{aL}\\
E_{aL}\end{pmatrix}\sim (1,3,-1/3+q/3),\\ 
&& e_{aR} \sim (1,1,-1),\hs E_{aR}\sim (1,1,q),\\
&& Q_{\al L} = \begin{pmatrix}
d_{\al L}\\
-u_{\al L}\\
J_{\al L} \end{pmatrix}\sim (3,3^*,-q/3),\\
&& Q_{3 L} = 
\begin{pmatrix}
u_{3L}\\
d_{3L}\\
J_{3L}\end{pmatrix}\sim (3,3,1/3+q/3),\\
&& u_{a R} \sim (3,1,2/3),\hs d_{aR}\sim (3,1,-1/3),\\
&& J_{\al R} \sim (3,1,-1/3 -q),\hs J_{3R}\sim (3,1,2/3 +q), \eea
where $a=1,2,3$ and $\al=1,2$ are family indices. The new fields $E_{a}$, $J_{\al}$, and $J_3$ have been introduced necessarily for completing the fermion representations and canceling all the anomalies. Notice that they possess electric charges, such as $Q(E_a)=q$, $Q(J_\al)=-1/3-q$, and $Q(J_3)=2/3+q$. If $q=1$, we achieve the 3-3-1 model with exotic charged leptons, and we denote $P_a\equiv E^+_a$ \cite{Pleitez:1992xh}. If $q=0$, the 3-3-1 model with neutral (heavy) fermions arises, and we define $N_a\equiv E^0_a$ \cite{Singer:1980sw}. For the typical versions, the minimal 3-3-1 model [the 3-3-1 model with right-handed neutrinos] are obtained by replacing $E_{aL}$ by $(e_{aR})^c$ [$(\nu_{aR})^c$, if imposed], whereas $E_{aR}$ is suppressed, which possess $q=1$ [$q=0$], respectively. This replacement does not apply for quarks, since the color and spacetime symmetries commute. Despite of the same $q$, the relevant models have alternative phenomenologies. Specially for $q=0$, $N_a$ may gain a charge $B-L=0$ different from that of $\nu_{aR}$, revealing a theory for dark matter \cite{Mizukoshi:2010ky,Dong:2013wca}. For $q=-1$, we obtain the 3-3-1 model with heavy charged leptons $(E^-_a)$ \cite{Pleitez:1994pu,Ozer:1995xi}. Interestingly, this version also implies dark matter stability if $E^-_{a}$ have $B-L=0$ different from that of the usual charged leptons (see, e.g., that in \cite{Dong:2015yra}). Along the line for $q=-1$, the flipped 3-3-1 model puts all quark families in antitriplets, while two lepton families in triplets and the remaining lepton family in sextet, which differs from the above arrangement \cite{Fonseca:2016tbn,Huong:2019vej}. This kind of the model has an extra chiral fermion triplet resided in the sextet, but it is highly degenerate in mass, negligibly contributing to $W$ mass. The other sources that affect $W$ mass are identical to the unflipped version with $q=-1$. That said, it is able to collect all the viable lepton sectors (family and left-handed indices omitted, right-handed counterparts if viable are in singlet) in Table \ref{331versions}, while the corresponding quark sectors are not listed, since they have a common form differing only in electric charge for exotic quarks.  

Three scalar triplets are generally introduced as
\bea 
\rho &=&
\begin{pmatrix}
\rho^+_1\\
\rho^0_2\\
\rho^{q+1}_3
\end{pmatrix}\sim (1,3,2/3+q/3),\\
\eta &=&
\begin{pmatrix}
\eta^0_1\\
\eta^{-}_2\\
\eta^q_3
\end{pmatrix}\sim (1,3,-1/3+q/3),\\
\chi &=&
\begin{pmatrix}
\chi^{-q}_1\\
\chi^{-q-1}_2\\
\chi^0_3
\end{pmatrix}\sim (1,3,-1/3-2q/3),\eea for which $\chi$ breaks the 3-3-1 symmetry down to the standard model, giving mass for new particles, while $\rho,\eta$ break the standard model symmetry down to $SU(3)_C\otimes U(1)_Q$, providing mass for ordinary particles. It is noted that one of the triplets $\rho,\eta$ may be excluded \cite{Dong:2014esa,Ferreira:2011hm}. By contrast, in flavor symmetry theories \cite{Dong:2010gk,Dong:2010zu,Dong:2011vb}, a large amount of scalar triplets may be introduced. Furthermore, scalar sextets may also be added to the present content \cite{Foot:1992rh,Dong:2008sw,Dong:2010gk,Dong:2010zu,Dong:2011vb}. However, the following investigation does not depend on such changes of scalar multiplet number; instead, it results from the vacuum structure of scalar fields. As shown in \cite{Dong:2005wgt}, it is sufficient to consider three triplets (given above) and one sextet, where in this work the sextet will be separately treated, without loss of generality. This vacuum structure depends on $q$ as well as $B-L$ behavior, studied below in order, along with implied gauge boson masses. 

Concerning $B-L$, it is stressed that $E_a,J_a$ in fermion multiplets generically have $B-L$ charge differently from that of the usual leptons and quarks, respectively. Let $[B-L](E_a)=n$. We obtain $B-L=\mathrm{diag}(-1,-1,n)$ for lepton triplets, which neither commutes nor closes algebraically with $SU(3)_L$. If $B-L$ is conserved, an extra $U(1)_{\mathcal{N}}$ group is required by symmetry principles such that $B-L=\beta' T_8+\mathcal{N}$, where $\beta'=-2(1+n)/\sqrt{3}$. We achieve $[B-L](J_3)=n+4/3$, $[B-L](J_\al)=-n-2/3$, and $[B-L](X,Y,\chi_{1,2})=-[B-L](\eta_3,\rho_3)=-n-1$, while the rest of fields takes usual value. Since $T_8$ is gauged, $B-L$ and $\mathcal{N}$ must be gauged. We impose $\nu_{aR}$ for cancelling $U(1)_{\mathcal{N}}$ anomalies and a superheavy scalar singlet $\phi_{B-L}$ with $B-L=2$ that couples to $\nu_{R}\nu_R$ and breaks $U(1)_{\mathcal{N}}$. After symmetry breaking, the neutrinos gain a small mass via canonical seesaw, while there exists a residual matter parity $P_M=(-1)^{3(B-L)+2s}$ not commuted with $SU(3)_L$ \cite{Dong:2013wca,Dong:2015yra}. The 3-3-1 model with $n=0$, as mentioned, possesses a nontrivial matter parity for new fields, $E_a$, $J_a$, $\eta_3$, $\rho_3$, $\chi_{1,2}$, $X$, and $Y$, such as 
\bea && P_M \left[\begin{array}{ccc} \nu_a &  d_\al & u_3  \\
e_a & -u_\al & d_3 \\
E_a & J_\al & J_3 \end{array}\right]=\left[\begin{array}{ccc} + & + & + \\
+ & + & + \\
- & -& - \end{array}\right],\\
&& P_M \left[\begin{array}{ccc}
\rho^+_1 & \eta^0_1 & \chi^{-q}_1 \\
\rho^0_2  & \eta^-_2  & \chi^{-q-1}_2 \\
\rho^{q+1}_3 & \eta^{q}_3 & \chi^{0}_3 \end{array}\right]=\left[\begin{array}{ccc} + & + & - \\
+ & + & - \\
- & -& + \end{array}\right],\\
&& P_M \left[\begin{array}{ccc}
(\ga,Z,Z') & W^+ & X^{-q} \\
W^- & (\ga,Z,Z')  & Y^{-q-1} \\
X^q & Y^{q+1} & (\ga,Z,Z') \end{array}\right]=\left[\begin{array}{ccc} + & + & - \\
+ & + & - \\
- & -& + \end{array}\right].\eea Hence, in spite of $q=0$ ($q=-1$), the relevant scalars $\eta_3,\chi_1$ ($\rho_3,\chi_2$) cannot develop a vacuum expectation value (VEV) due to the matter parity conservation. Alternatively, the 3-3-1 model with $n=1$, including the minimal 3-3-1 model, the 3-3-1 model with right-handed neutrinos, and even the 3-3-1 model with heavy charged leptons, transform trivially under $P_M$, i.e. $P_M=1$ for every field. In this case, the scalars $\eta_3,\chi_1$ ($\rho_3,\chi_2$) if electrically neutral can develop a VEV, not protected by $P_M$, actually governed by $B-L$ (or $P_M$) violating interactions. With the aid of $P_M$, a summary of possible vacuum structures for 3-3-1 variants is given in Table \ref{331versions}. 

If the $B-L$ symmetry is approximate, we avoid a $U(1)_{\mathcal{N}}$ extension as given above. In this case, the interactions violating $B-L$ enter, such as \bea && s^u_{3a} \bar{Q}_{3L}\chi u_{aR}+s^d_{\al a}\bar{Q}_{\al L}\chi^* d_{aR} +s^J_{33} \bar{Q}_{3L}\eta J_{3R}+s^J_{\al \beta}\bar{Q}_{\al L}\eta^* J_{\beta R}+s^J_{3\al}\bar{Q}_{3L}\rho J_{\al R} + s^J_{\al 3} \bar{Q}_{\al L}\rho^* J_{3R}\crn
&&+\bar{\mu}^2\eta^\dagger \chi + (\bar{\la}_1\eta^\dagger \chi + \bar{\la}_2 \eta^\dagger \eta +\bar{\la}_3 \rho^\dagger \rho +\bar{\la}_4 \chi^\dagger \chi)\eta^\dagger \chi +\bar{\la}_5 (\eta^\dagger \rho) (\rho^\dagger \chi)+H.c.,\eea for the 3-3-1 model with right-handed neutrinos ($q=0,n=1$), besides the normal Yukawa couplings ($h$'s) and scalar self-couplings ($\mu$'s and $\la$'s)---as in the usual theory---which conserve $B-L$. It is easily verified that all these violating couplings violate $L$ by 2 units, except for $\bar{\la}_1$ by 4 units, while preserve $B$ and a lepton parity $(-1)^L$. Thus, the proton stability is protected by $B$ and $(-1)^L$. Furthermore, the violating couplings must be small, e.g. $\bar{\mu}\ll \mu$, $\bar{\la}\ll \la$, and $s\ll h$, since by contrast $L$-conservation sets $\bar{\mu},\bar{\la},s=0$ but $\mu,\la,h \neq 0$. In this case, the potential minimization gives $L$-violating VEVs, $\langle \eta_3,\chi_1 \rangle \sim \bar{\mu}^2/w$, as suppressed, where $w$ is the 3-3-1 breaking scale (cf. \cite{VanLoi:2019eax}). If discarding the unwanted coupling $\psi_L\psi_L\rho$ by a symmetry $\rho\to -\rho$, the neutrinos gain a naturally small mass via $L$-violating effective interaction, $\fr{s'}{w}\psi_L \psi_L \eta\eta$, to be $m_\nu \sim s' u^2/w$, doubly suppressed by $s'\ll h$ and $u\ll w$, where $u$ is a weak scale. The presence of small $\langle \eta_3,\chi_1 \rangle$ and $s$'s lead to a small mixing between usual quarks and exotic quarks causing flavor changing $Z$-currents, as studied below (Sec.~\ref{Constrain}).   

\subsection{Gauge spectrum for $q\neq 0,-1$ or $B-L$ conservation}

For $q\neq 0$ and $q\neq -1$, the scalar components that are electrically neutral can develop VEVs, such as   
\be \langle \rho\rangle =
\fr{1}{\sqrt{2}} \left(
\begin{array}{c}
0\\
v \\
0
\end{array}\right),\hs \langle \eta \rangle = \fr{1}{\sqrt{2}}\left(
\begin{array}{c}
u \\
0\\
0
\end{array}\right),\hs
\langle \chi\rangle =
\fr{1}{\sqrt{2}} \left(
\begin{array}{c}
0\\
0\\
w
\end{array}\right).\label{vev1}\ee  
Since $w$ breaks the 3-3-1 symmetry, while $u,v$ break the standard model symmetry, we impose $w\gg v,u$ for consistency. This standard vacuum alignment has been extensively studied in the literature, even applying for the 3-3-1 model with arbitrary $q$ (see Table \ref{331versions}). However, notice that for the model with $q=0$ ($q=-1$), an extra symmetry such as $B-L$ and its residual matter parity is needed to prevent the {\it other neutral} scalars $\eta_3,\chi_1$ ($\rho_3,\chi_2$) from developing a VEV, ensuring the standard vacuum structure, as mentioned \cite{Dong:2013wca,Dong:2015yra}. Particularly for $q=0$, this section applies for the 3-3-1 model with neutral (heavy) fermions, not for the 3-3-1 model with right-handed neutrinos.       

The mass spectrum of the gauge bosons is given by \be \mathcal{L} \supset \sum_{\Phi}(D^\mu\langle \Phi\rangle)^\dagger (D_\mu \langle \Phi\rangle),\ee where $\Phi$ runs over the scalar triplets, and the covariant derivative is $D_\mu = \pa_\mu + i g_s t_j G_{j\mu} + i g T_j A_{j\mu} + i g_X X B_\mu$. Here ($g_s,g,g_X$) and ($G_j,A_j,B$) are the gauge couplings and gauge bosons of 3-3-1 subgroups, respectively, and $t_j$ is $SU(3)_C$ charges. 

Define non-Hermitian gauge bosons,
\be W^\pm = \fr{A_1\mp i A_2}{\sqrt{2}},\hs X^{\mp q} = \fr{A_4\mp i A_5}{\sqrt{2}},\hs Y^{\mp (1+q)} = \fr{A_6\mp i A_7}{\sqrt{2}},\ee which are coupled to the weight-raising and -lowering operators, \be T_\pm=\fr{T_1\pm i T_2}{\sqrt{2}},\hs U_\pm=\fr{T_4\pm i T_5}{\sqrt{2}},\hs  V_\pm=\fr{T_6\pm i T_7}{\sqrt{2}},\ee respectively. $W^\pm$, $X^{\mp q}$, and $Y^{\mp (1+q)}$ are physical fields by themselves with masses
\be m^2_{W} = \fr{g^2}{4}(v^2+u^2),\hs m^2_X=\fr{g^2}{4}(w^2+u^2),\hs m^2_Y=\fr{g^2}{4}(w^2+v^2),\label{wxymass} \ee respectively.  
$W$ is identical to that of the standard model, while $(X,Y)$ form a new, heavy gauge vector doublet with a mass splitting $|m^2_Y-m^2_X|< m^2_W$. 

Consider neutral gauge bosons. The photon field $A$ that is coupled to the electric charge $Q$ with coupling $e$ is given by $A/e=(A_3+\beta A_8)/g+B/g_X$, which is determined from $Q=T_3+\beta T_8+X$ by substituting each generator with corresponding gauge field over coupling. This is a direct result of electric charge conservation, neither depending on VEVs nor necessarily diagonalizing the relevant mass matrix, as proved in \cite{Dong:2005wgt}. The normalization of photon field implies $s_W\equiv e/g=g_X/\sqrt{g^2+(1+\beta^2)g_X^2}$, which matches the sine of the Weinberg angle in the standard model. Hence, the photon field $A$ can be rewritten as  
\be A = s_W A_3 + c_W \left(\beta t_W A_8 +\sqrt{1-\beta^2 t^2_W}B\right),\ee where the expression in the parentheses is just the hypercharge field defined by (\ref{gt3}). The standard model $Z$ field is given orthogonally to $A$, 
\be Z = c_W A_3 - s_W \left(\beta t_W A_8 +\sqrt{1-\beta^2 t^2_W}B\right),\ee as usual, while a new $Z'$ field is obtained orthogonally to the hypercharge field,
\be Z' = \sqrt{1-\beta^2 t^2_W} A_8 - \beta t_W B.\ee   
 
The photon $A$ is massless and decoupled, as a physical field, while $Z$ and $Z'$ mix via a symmetric mass matrix with elements, given by
 \bea
 m^2_Z&=&\fr {g^2}{4c^2_W} (v^2+u^2),\\
 m^2_{ZZ'} &=& \fr{g^2\left[(\sqrt3\beta t^2_W-1)v^2+(\sqrt3\beta t^2_W+1)u^2\right]}{4\sqrt3 c_W\sqrt{1-\beta^2 t^2_W}},\\ 
 m^2_{Z'}&= &\fr{g^2\left[4w^2+(\sqrt3\beta t^2_W-1)^2v^2+(\sqrt3\beta t^2_W+1)^2u^2\right]}{12(1-\beta^2 t^2_W)}.\eea Diagonalizing the $Z$-$Z'$ mass matrix, we obtain two physical fields, 
\be Z_1=c_\varphi Z - s_\varphi Z',\hs  Z_2=s_\varphi Z + c_\varphi Z',\ee
where the $Z$-$Z'$ mixing angle ($\varphi$) and $Z_{1,2}$ masses are 
  \bea 
     t_{2\varphi} &=& \frac{2m^2_{ZZ'}}{m^2_{Z'}-m^2_Z}\simeq \fr{\sqrt{3(1-\beta^2 t^2_W)}\left[(\sqrt3\beta t^2_W-1)v^2+(\sqrt3\beta t^2_W+1)u^2\right]}{2w^2 c_W},\\
     m^2_ {Z_1} &=& \fr1 2 \left[m^2_Z +m^2_{Z'} - \sqrt{(m^2_Z - m^2_{Z'})^2 + 4m^4_ {ZZ'} }\right]
     \simeq m^2_Z-\frac{m^4_{ZZ'}}{m^2_{Z'}}\crn
     &\simeq& \fr{g^2}{4c^2_W}\left\{v^2+u^2-\fr{\left[(\sqrt3\beta t^2_W-1)v^2+(\sqrt3\beta t^2_W+1)u^2\right]^2}{4w^2}\right\},\\
    m^2_ {Z_2} &=& \fr1 2 \left[m^2_Z +m^2_{Z'} + \sqrt{(m^2_Z - m^2_{Z'})^2 + 4m^4_ {ZZ'} }\right]\simeq m^2_{Z'}.
 \eea
The mixing angle $\varphi$ is small, suppressed by $(v,u)^2/w^2$. Additionally, $Z_1$ has a mass approximating that of $Z$, called the standard model $Z$-like boson, while $Z_2$ is a new, heavy gauge boson with mass proportional to $w$. 

\subsection{\label{iicdt}Gauge spectrum for $q= 0,-1$ and $B-L$ violation}

For $q=0$ ($q=-1$), the other scalars $\eta_3,\chi_1$ ($\rho_3,\chi_2$) are electrically neutral and possibly develop a VEV in addition to the VEVs $u,v,w$, given above \cite{Dong:2006mg,Boucenna:2014ela,Dong:2017ayu}. In this case, the $B-L$ symmetry as mentioned is necessarily violated \cite{Dong:2013ioa,Dong:2014esa}, since $\eta_3,\chi_1$ ($\rho_3,\chi_2$) have nonzero $B-L$ number. This yields a vacuum structure, called abnormal vacuum, different from the previous model (see Table \ref{331versions}). Particularly for $q=0$, this section applies for the 3-3-1 model with right-handed neutrinos, not for the 3-3-1 model with neutral (heavy) fermions. Because the physics obtained is the same independent of $q=0$ or $-1$ for phenomena interested in this work, we consider only $q=0$, thus $\beta=-1/\sqrt{3}$. In this case, although both $\eta$ and $\chi$ transform the same under the gauge symmetry, they differ in $B-L$ numbers, possibly obtaining VEVs at the first and third components. That said, the vacuum alignment  under consideration is generically given by
\be \langle \rho\rangle =
\fr{1}{\sqrt{2}} \left(
\begin{array}{c}
0\\
v \\
0
\end{array}\right),\hs\langle \eta \rangle = \fr{1}{\sqrt{2}}\left(
\begin{array}{c}
u \\
0\\
w'
\end{array}\right),\hs
\langle \chi\rangle =
\fr{1}{\sqrt{2}} \left(
\begin{array}{c}
u' \\
0\\
w
\end{array}\right).\label{vev2}\ee 
In contrast to $u$, $v$, and $w$ that conserve $B-L$, the remaining VEVs $u',w'$ break this number, suppressed by relevant violation interactions. To be consistent with the standard model, we impose $u'\ll u$ and $w'\ll w$, in addition to $u,v\ll w$ \cite{Dong:2008sw}.
   
Substituting the VEVs in (\ref{vev2}) to the Lagrangian, we get
\bea
\mathcal{L} &\supset &\begin{pmatrix} W^- & Y^- \end{pmatrix} M^2_c \begin{pmatrix} W^+ &Y^+ \end{pmatrix}^T+\frac{g^2}{8}(w^2+u^2+w'^2+u'^2)A^2_5\crn
&&+\frac{1}{2}\begin{pmatrix}A_3&A_8&B&A_4\end{pmatrix}M^2_0 \begin{pmatrix}A_3&A_8&B&A_4\end{pmatrix}^T,
\eea 
where $W^{\pm}=(A_1\mp iA_2)/\sqrt2$ and $Y^{\pm}=(A_6\pm iA_7)/\sqrt2$ defined as before mix, while the real and imaginary parts of $X^{0,0*}=(A_4\mp i A_5)/\sqrt{2}$ behave differently, i.e. $A_4$ mixes with $A_{3,8}$ and $B$, whereas $A_5$ does not. That said, $A_5$ is decoupled, as physical field, with mass,
\be m^2_{A_5} = \frac{g^2}{4}(w^2+u^2+w'^2+u'^2). \ee The mass matrices of the charged and neutral gauge bosons are given, respectively, by
\bea M^2_c &=& \frac{g^2}{4}\left(\begin{array}{cc}v^2+u^2+u'^2 & w u'+uw' \\ wu'+uw' & w^2+v^2+w'^2\end{array} \right),\\ 
 M^2_0 &=& \frac{g^2}{4}\left(\begin{array}{cccc} v^2+u^2+u'^2 & \frac{u^2-v^2+u'^2}{\sqrt3} & -\frac{2(2v^2+u^2+u'^2)t_X}{3} & wu'+uw' \\
\frac{u^2-v^2+u'^2}{\sqrt3} & \frac{4w^2+v^2+u^2+4w'^2+u'^2}{3} & \frac{2[2(w^2+v^2+w'^2)-u^2-u'^2]t_X}{3\sqrt3} & -\frac{wu'+uw'}{\sqrt3}\\
-\frac{2(2v^2+u^2+u'^2)t_X}{3} & \frac{2[2(w^2+v^2+w'^2)-u^2-u'^2]t_X}{3\sqrt3} & \frac{4(w^2+4v^2+u^2+w'^2+u'^2)t_X^2}{9} & -\frac{4(wu'+uw')t_X}{3}\\
wu'+uw' & -\frac{wu'+uw'}{\sqrt3} & -\frac{4(wu'+uw')t_X}{3} & \frac{4}{g^2}m^2_{A_5}
\end{array} \right),\eea
where $t_X=g_X/g$. It is noteworthy that both the mixing of $W$ and $Y$ and the mixing of ($A_3$, $A_8$, $B$) and $A_4$ are caused by $u',w'$. 

Diagonalizing $M^2_c$, we obtain two physical fields
\be W_1=c_\theta W - s_\theta Y,\hs  Y_1 = s_\theta W + c_\theta Y,\ee
where the $W$-$Y$ mixing angle ($\theta$) is given by
\be t_{2\theta} = \frac{2(w u'+uw')}{w^2-u^2+w'^2-u'^2},\ee
which implies that $\theta\simeq u'/w+(u/w)(w'/w)$ is very small, because of $u'\ll u$ and $u,w'\ll w$. The $W_1,Y_1$ masses are 
\bea m^2_{W_1} &=& \frac{g^2}{8}\left[w^2+2v^2+u^2+w'^2+u'^2-\sqrt{(w^2-u^2+w'^2-u'^2)^2+4(wu'+uw')^2}\right]\crn
&\simeq& \fr{g^2}{4}\left(v^2+u^2-\fr{2uu'w'}{w}-\fr{u^2w'^2}{w^2}\right),\\
m^2_{Y_1} &=& \frac{g^2}{8}\left[w^2+2v^2+u^2+w'^2+u'^2+\sqrt{(w^2-u^2+w'^2-u'^2)^2+4(wu'+uw')^2}\right]\crn
&\simeq&  \fr{g^2}{4}\left(w^2+v^2+w'^2+u'^2\right).
 \eea
$W_1$ is the standard model $W$-like boson, whereas $Y_1$ is a new, heavy charged gauge boson.

To diagonalize $M^2_0$, we define the photon $A$, the usual $Z$, and the new $Z'$ as in the previous model, but for $\beta=-1/\sqrt{3}$, such as
\bea A &=& s_W A_3 + c_W \left(-\fr{t_W}{\sqrt3} A_8 +\sqrt{1-\fr{t^2_W}{3}}B\right),\\
Z &=& c_W A_3 - s_W \left(-\fr{t_W}{\sqrt3} A_8 +\sqrt{1-\fr{t^2_W}{3}}B\right),\\
Z^\prime &=& \sqrt{1-\fr{t^2_W}{3}} A_8 + \fr{t_W}{\sqrt3} B,\eea  
where $s_W=\sqrt3 g_X/\sqrt{3g^2+4g_X^2}$. In the new basis ($A$, $Z$, $Z'$, $A_4$), the photon $A$ is massless and decoupled, as usual, while $Z$, $Z'$, and $A_4$ mix by themselves via the mass matrix,
\be M^{\prime 2}_0 = \frac{g^2}{4}\left(\begin{array}{ccc} \frac{v^2+u^2+u'^2}{c^2_W} & \frac{(u^2+u'^2)c_{2W}-v^2}{c^2_W\sqrt{1+2c_{2W}}} & \frac{wu'+uw'}{c_W}\\
\frac{(u^2+u'^2)c_{2W}-v^2}{c^2_W\sqrt{1+2c_{2W}}} & \frac{4(w^2+w'^2)c^4_W+v^2+(u^2+u'^2)c^2_{2W}}{c^2_W(1+2c_{2W})} & -\frac{wu'+uw'}{c_W\sqrt{1+2c_{2W}}}\\
\frac{wu'+uw'}{c_W} & -\frac{wu'+uw'}{c_W\sqrt{1+2c_{2W}}} & \frac{4}{g^2}m^2_{A_5}
\end{array} \right). \ee

It is easily verified that $M^{\prime 2}_0$ (thus $M^2_0$) contains an exact eigenvalue,
\be m^2_{\mathcal{A}_4} = \frac{g^2}{4}(w^2+u^2+w'^2+u'^2), \ee
with a corresponding exact eigenstate,
\be \mathcal{A}_4
= s_{\theta'}Z+c_{\theta'}\left(t_{\theta'}\sqrt{3-4s^2_{W}}Z'+\sqrt{1-t^2_{\theta'}(3-4s^2_{W})}A_4\right), \ee
where $s_{\theta'} = s_{2\theta}/2c_W$ is very small as $\theta$ is, hence $\mathcal{A}_4\simeq A_4$. It is noteworthy that $\mathcal{A}_4$ and $A_5$ always have equal masses.\footnote{This occurs for generic vacuum structure---including $u' \neq 0$ and $w'\neq 0$---that conserves the electric charge in spite of the fact that $A_4$ mixes with $Z,Z'$.} Hence, we identify 
\be X^{0,0*}_1 =\frac{\mathcal{A}_4\mp iA_5}{\sqrt2} \ee
to be a physical non-Hermitian field, with the common mass, \be m^2_{X_1}=\frac{g^2}{4}(w^2+u^2+w'^2+u'^2).\ee 

To diagonalize $M^{\prime 2}_0$, we choose two gauge vectors orthogonally to $\mathcal{A}_4$, such as  
\bea \mathcal{Z} &=& c_{\theta'}Z-s_{\theta'}\left(t_{\theta'}\sqrt{3-4s^2_{W}}Z'+\sqrt{1-t^2_{\theta'}(3-4s^2_{W})}A_4\right),\\
\mathcal{Z'} &=& \sqrt{1-t^2_{\theta'}(3-4s^2_{W})}Z'-t_{\theta'}\sqrt{3-4s^2_{W}}A_4,\eea where $\mathcal{Z}\simeq Z$ and $\mathcal{Z}'\simeq Z'$ similar to $\mathcal{A}_4\simeq A_4$, since $|\theta'|\ll 1$. In the new basis ($\mathcal{Z}$, $\mathcal{Z'}$, $\mathcal{A}_4$), the field $\mathcal{A}_4$ is decoupled, while $\mathcal{Z}$ and $\mathcal{Z'}$ mix by themselves via a symmetric mass matrix with elements, given by
\bea m^2_{\mathcal{Z}} &=& \frac{g^2[4v^2+(u^2+u'^2)(3+c^2_{2\theta})-(w^2+w'^2)s^2_{2\theta}]}{4(4c^2_W-s_{2\theta}^2)},\\
m^2_{\mathcal{Z}\mathcal{Z'}} &=& \frac{g^2\{[(u^2+u'^2)c_{2W}-v^2]c_{2\theta}-(wu'+uw')(1+2c_{2W})s_\theta c_\theta\}}{\sqrt{1+2c_{2W}}(4c^2_W-s_{2\theta}^2)},\\
m^2_{\mathcal{Z'}} &=& \frac{g^2[(w^2+w'^2)(16c^4_W-s^2_{2\theta})+4v^2c^2_{2\theta}+(u^2+u'^2)(4c^2_{2W}-s^2_{2\theta})]}{4(1+2c_{2W})(4c^2_W-s^2_{2\theta})}.\eea
Diagonalizing this mass matrix, we obtain physical fields,
\be Z_1=c_\varphi\mathcal{Z} - s_\varphi\mathcal{Z'},\hs  Z_2=s_\varphi\mathcal{Z} + c_\varphi\mathcal{Z'},\ee
where the $\mathcal{Z}$-$\mathcal{Z'}$ mixing angle ($\varphi$) and $Z_1,Z_2$ masses are
\bea 
   t_{2\varphi} &=& \frac{2m^2_{\mathcal{Z}\mathcal{Z'}}}{m^2_{\mathcal{Z'}}-m^2_{\mathcal{Z}}}\simeq \frac{\sqrt{3-4s^2_W}(u^2c_{2W}-v^2)}{2w^2c^4_W},\\
    m^2_ {Z_1} &=& \fr1 2 \left[m^2_{\mathcal{Z}} +m^2_{\mathcal{Z'}} - \sqrt{(m^2_{\mathcal{Z}} - m^2_{\mathcal{Z'}})^2 + 4m^4_ {\mathcal{Z}\mathcal{Z'}} }\right]\simeq m^2_\mathcal{Z}-\frac{m^4_{\mathcal{Z}\mathcal{Z}'}}{m^2_{\mathcal{Z}'}}\crn
    &\simeq& \fr{g^2}{4c^2_W}\left[v^2+u^2-\fr{2uu'w'}{w}-\fr{u^2w'^2}{w^2}-\fr{(u^2c_{2W}-v^2)^2}{4c^4_W w^2}\right],\\
   m^2_ {Z_2} &=& \fr1 2 \left[m^2_{\mathcal{Z}} +m^2_{\mathcal{Z'}} + \sqrt{(m^2_{\mathcal{Z}} - m^2_{\mathcal{Z'}})^2 + 4m^4_ {\mathcal{Z}\mathcal{Z'}} }\right]\simeq m^2_{\mathcal{Z}'}. 
 \eea
The $\varphi$ angle is small, suppressed by $(u,v)^2/w^2$. Additionally, the field $Z_1$ is the standard model $Z$-like boson, while $Z_2$ is a new, heavy gauge boson with mass at $w$.

\section{\label{Wmass}Sources of the $W$-mass shift}

The $W$-mass shift in the 3-3-1 model arises from various new physics sources, $Z$-$Z'$ mixing, $X$-$Y$ and $Z$-$Z'$-$X$ mixings, a gauge vector doublet as well as a new Higgs doublet that are not degenerate in mass. They are newly recognized, depending on relevant 3-3-1 model.  

Let us remind the reader that as already done in the literature of electroweak precision fit \cite{Lu:2022bgw,Strumia:2022qkt,Cacciapaglia:2022xih,Asadi:2022xiy,Gu:2022htv,Paul:2022dds,Balkin:2022glu,Carpenter:2022oyg,Bagnaschi:2022whn}, a positive and dominant contribution of the Peskin-Takeuchi $T$-parameter can generate an enhancement of the $W$-boson mass consistent with the recent CDF measurement.     

\subsection{$Z$-$Z'$ mixing in the model with $q\neq 0,-1$ or $B-L$ conservation}

The 3-3-1 model under consideration reveals a tree-level mixing of $Z$ with $Z'$, while $W$ is retained. Because of the $Z$-$Z'$ mixing, the observed $Z_1$-boson mass ($m_{Z_1}$) is reduced in comparison with the standard model $Z$-boson mass ($m_Z$). This gives rise to a positive contribution to the $T$ parameter at tree level, 
\be \al T=\rho-1=\fr{m^2_W}{c^2_W m^2_{Z_1}}-1=\fr{m^2_Z}{m^2_{Z_1}}-1\simeq \fr{\left[(\sqrt3\beta t^2_W-1)v^2+(\sqrt3\beta t^2_W+1)u^2\right]^2}{4w^2(v^2+u^2)}, \ee where note that $m_W=m_Z c_W$ \cite{Holdom:1990xp}. Since $m_{Z_1}$ is precisely measured and fixed, this enhances the mass of $W$ boson proportionally to $\al T$, such as  \cite{Peskin:1990zt,Peskin:1991sw,Maksymyk:1993zm}
\be \Delta m^2_W = \frac{c^4_W m^2_Z}{c^2_W-s^2_W}\left(\fr{m^2_Z}{m_{Z_1}^2}-1\right)\simeq \frac{c^4_W m^2_Z}{c^2_W-s^2_W}\fr{\left[(\sqrt3\beta t^2_W-1)v^2+(\sqrt3\beta t^2_W+1)u^2\right]^2}{4w^2(v^2+u^2)}. \ee

\begin{figure}[!h]
\begin{center}
	\includegraphics[scale=0.28]{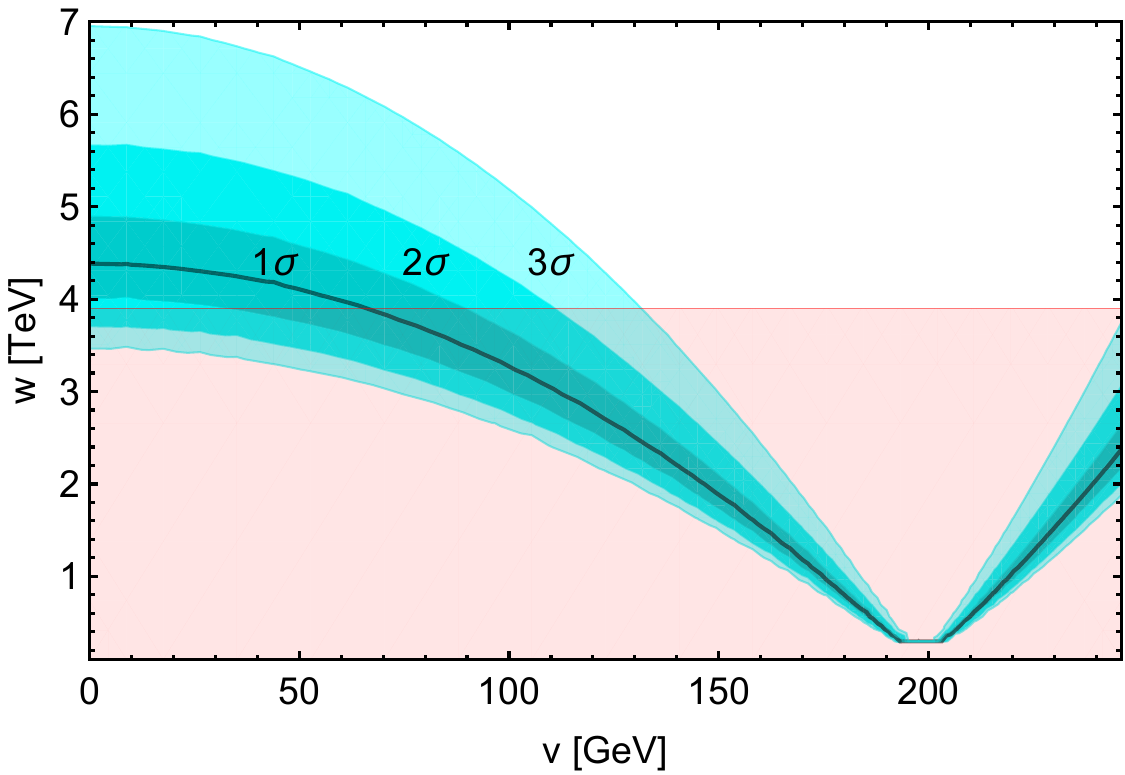}
	\includegraphics[scale=0.28]{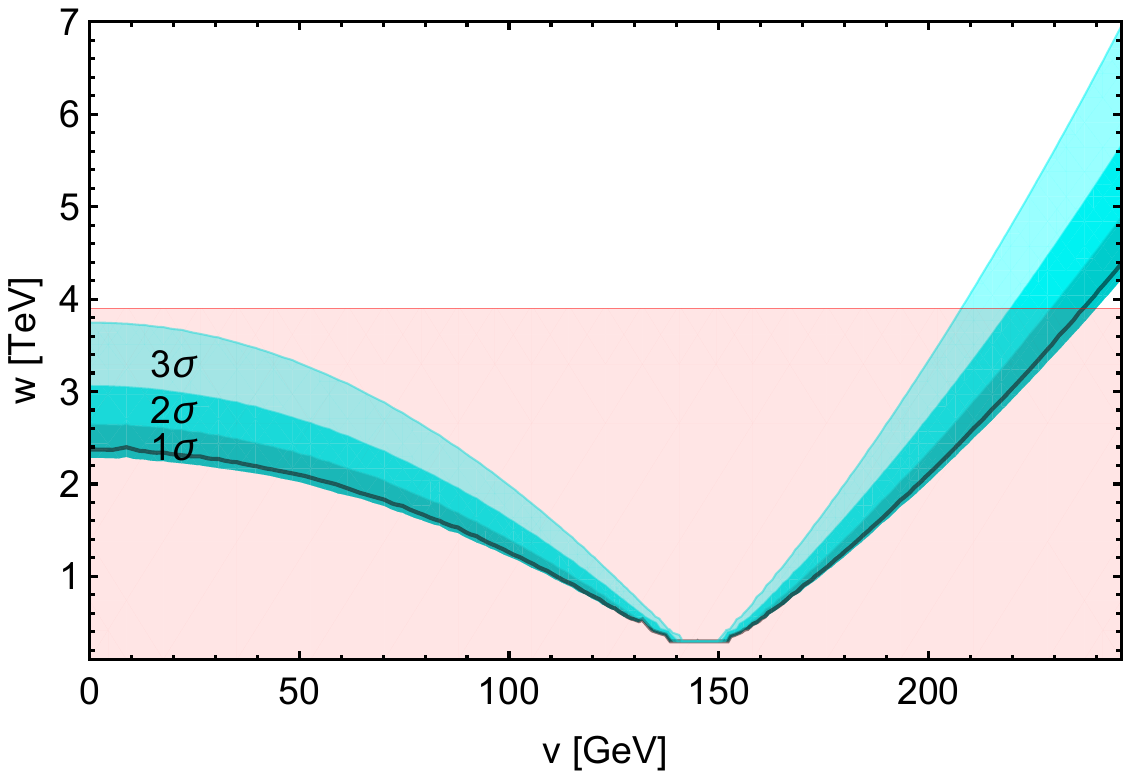}
	\includegraphics[scale=0.28]{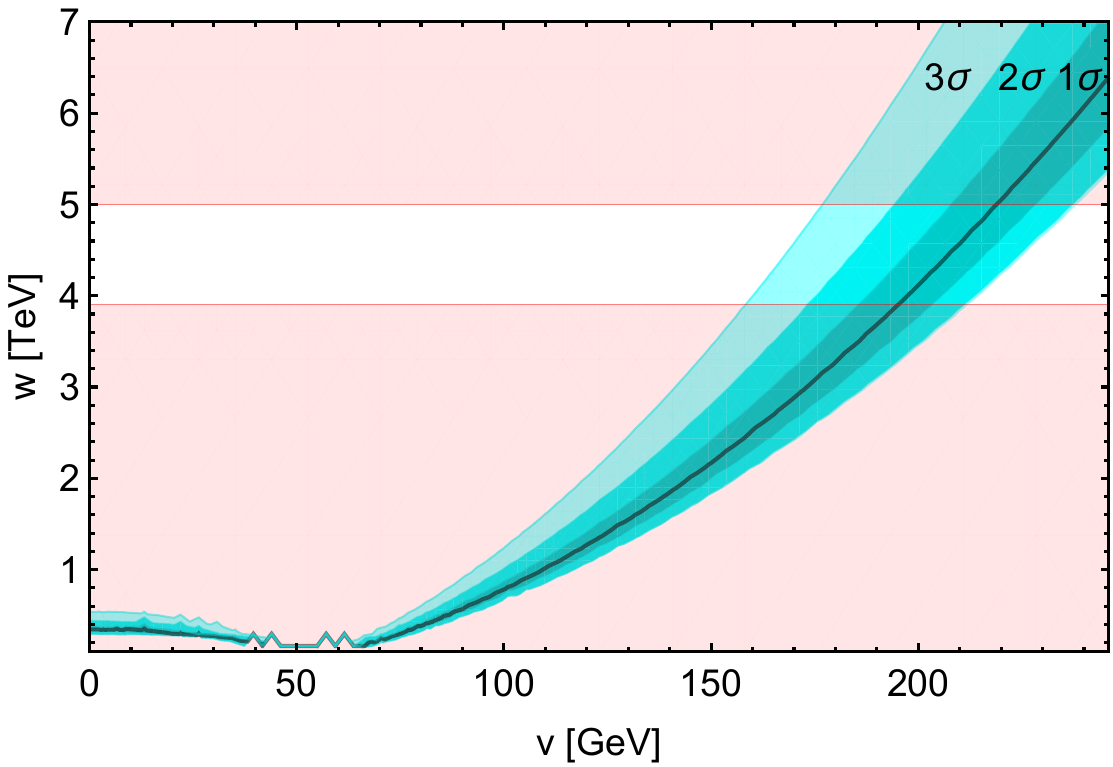}
		\caption{\label{CDF1} Viable ($v,w$) regimes bounded by the CDF $W$-boson mass at 1$\sigma$, 2$\sigma$, and $3\sigma$ ranges (in cyans), according to the 3-3-1 model with $\beta=1/\sqrt3$ (left panel), $\beta=-1/\sqrt3$ (middle panel), and $\beta=-\sqrt3$ (right panel), where the FCNC (lower light-red) and Landau pole (upper light-red, if applied) bounds are also included.}	
	\end{center}
\end{figure}

We input the parameters as $\al\simeq 1/127.955$, $s_W^2\simeq 0.231$, and $u^2+v^2=246^2$ GeV$^2$. Additionally, we consider the three 3-3-1 models according to $\beta=1/\sqrt{3}$, $\beta=-1/\sqrt{3}$, and $\beta=-\sqrt{3}$, where the last two are identical to the 3-3-1 model with neutral (heavy) fermions and the minimal 3-3-1 model, respectively. We make a contour of $\Delta m^2_{W} = 80.4335^2 - 80.357^2$ GeV$^2$ taking central values \cite{CDF:2022hxs,ParticleDataGroup:2020ssz} as function of $v$ and $w$ for the mentioned models as in Fig.~\ref{CDF1}. Here, the black line is for the central value. Additionally, the $1\sigma, \, 2\sigma$, and $3\sigma$ ranges are also shown (in cyans). Besides, the excluded region (light red) by the FCNCs and the Landau pole limit if it applies (taken as $5$ TeV) have appropriately been included to plot (cf. Sec.~\ref{Constrain}). 

From Fig. \ref{CDF1}, we obtain the viable ranges for the new physics and weak scales, namely, ($3.9$ TeV $<w<4.4$ TeV and $0$ GeV $<v<65.7$ GeV), ($3.9$ TeV $<w<4.4$ TeV and $237.1$ GeV $<v<246$ GeV), and ($3.9$ TeV $<w<5$ TeV and $195.2$ GeV $<v<219.1$ GeV) according to the 3-3-1 model with $\beta=1/\sqrt3$, $\beta=-1/\sqrt3$, and $\beta=-\sqrt3$, respectively. Note that the weak scale $u$ depends on $v$ as $u=\sqrt{(246\text{ GeV})^2-v^2}$.

\subsection{$W$-$Y$ and $Z$-$Z'$-$X$ mixings in the model with $q=0$ and $B-L$ violation}

It is clear that when $q=0$, thus $\beta=-1/\sqrt3$, the scalar triplets, $\eta$ and $\chi$, in the 3-3-1 model with right-handed neutrinos have two electrically-neutral entries at the top and bottom and they can all develop VEVs. However, the VEVs $u'$,$w'$ violate lepton number and should be small, i.e. $u'\ll u$ and $w'\ll w$, as given. These $u',w'$ cause the mixings between the usual $W$ boson and new charged gauge boson $Y$ as well as among $Z,Z'$ and new neutral gauge boson $X$. Whereas, the normal VEVs $u,v,w$ induce $Z$--$Z'$ mixing similarly to the previous model. 

Because of the mixings, both physical masses of $W$ and $Z$ bosons, i.e. $W_1$ and $Z_1$ respectively, are reduced in comparison with the standard model values, such as
\bea m^2_W &=& \fr{g^2}{4}(v^2+u^2+u'^2)\to m^2_{W_1}\simeq \fr{g^2}{4}\left(v^2+u^2-\fr{2uu'w'}{w}-\fr{u^2w'^2}{w^2}\right),\label{W1mass}\\
m^2_Z &=& \fr{g^2}{4c^2_W}(v^2+u^2+u'^2)\to m^2_{Z_1}\simeq \fr{g^2}{4c^2_W}\left[v^2+u^2-\fr{2uu'w'}{w}-\fr{u^2w'^2}{w^2}-\fr{(u^2c_{2W}-v^2)^2}{4c^4_W w^2}\right],\label{Z1mass}\eea a phenomenon occurs similarly to the economical 3-3-1 model \cite{Dong:2006mg}. New observation is that the reduction of $Z$ mass is bigger than that of $W$ mass. Correspondingly, this produces a $\rho$-parameter bigger than 1, causing the CDF $W$-mass shift, as measured. 

That said, from (\ref{W1mass}) and (\ref{Z1mass}) we obtain a positive contribution to the $T$ parameter at tree level, such as 
\be \al T=\rho-1=\frac{m^2_{W_1}}{c^2_W m^2_{Z_1}}-1\simeq \fr{(u^2c_{2W}-v^2)^2}{4c^4_W(u^2+v^2)w^2}. \ee It is noteworthy that the three terms that do change the $W,Z$ masses in (\ref{W1mass}) and (\ref{Z1mass}) are at the same order, because of $u'/u\sim w'/w\sim (u,v)/w$. Additionally, the first two of these terms caused by $u',w'$ come from $W$--$Y$ and $(Z,Z')$--$X$ mixings, while the last term suppressed by $(u,v)^2/w^2$ arises from the $Z$--$Z'$ mixing. Although both kinds of the mixings reduce the $W,Z$ masses, only the $Z$--$Z'$ mixing governs the $\rho$-parameter deviation which subsequently affects the $W$ mass as shifted, similar to the previous model. 

That said, the enhancement of the $W$-boson mass is simply given by \cite{Peskin:1990zt,Peskin:1991sw,Maksymyk:1993zm},
\be \Delta m^2_W = \frac{c^4_W m^2_Z}{c^2_W-s^2_W}\left(\frac{m^2_{W_1}}{c^2_W m^2_{Z_1}}-1\right)\simeq  \frac{c^4_W m^2_Z}{c^2_W-s^2_W}\fr{(u^2c_{2W}-v^2)^2}{4c^4_W(u^2+v^2)w^2}.\ee The result in the previous sector for $\beta=-1/\sqrt{3}$ applies to this model without change, however.  Although the phenomenologies of the two mentioned models are distinct, characterized by $B-L$ conservation or violation, the $Z$--$Z'$ mixing is crucial to set the CDF $W$-mass anomaly. The mixing effects caused by $u',w'$ effectively not contributing to the $W$-mass shift as observed are probably due to the fact that $u',w'$ break lepton number, associated with Majoron fields that are eaten by the corresponding non-Hermitian gauge bosons $(X,Y)$. Hence, $u',w'$ only affect the $X,Y$ observables.

\subsection{Oblique contributions of the non-degenerate vector doublet}

At one-loop level, the two new non-Hermitian gauge bosons predicted by the 3-3-1 model $X^{\pm q}$ and $Y^{\pm(1+q)}$, which form an $SU(2)_L$ doublet according to the decomposition of $SU(3)_L$ gauge adjoint $8=3\oplus 2\oplus 2^*\oplus 1$, contribute to the oblique parameters $S$, $T$, and $U$, through transverse self-energies. The $W$-boson mass shift induced by these oblique corrections can be expressed as follows \cite{Peskin:1990zt,Peskin:1991sw,Maksymyk:1993zm}
\be \Delta m^2_{W} = \frac{c_W^2 m_Z^2\al}{c_W^2-s_W^2}\left(-\frac{S}{2}+c_W^2 T+\frac{c_W^2-s_W^2}{4s_W^2}U\right). \label{WmassSTU}\ee

For the 3-3-1 model with right-handed neutrinos, i.e. $\beta=-1/\sqrt3$, the oblique corrections have already been computed, given by \cite{Long:1999bny},
\bea S &=& \frac{1}{4\pi}\left[5\ln\frac{m^2_Y}{m^2_X}+\left(5+ \frac{4m^2_Y}{3m^2_Z}\right)\bar{F}_0(m^2_Z,m_Y,m_Y)-\left(1+\frac{4m^2_X}{3m^2_Z}\right)\bar{F}_0(m^2_Z,m_X,m_X) \right],\label{SRHN}\\
T &=& \fr{3\sqrt{2}G_F}{16\pi^2\al}\left(m^2_Y+m^2_X-\fr{2m^2_Y m^2_X}{m^2_Y-m^2_X}\ln\fr{m^2_Y}{m^2_X}\right)\crn
&&+\fr{1}{4\pi s^2_W}\left(\fr{m^2_Y+m^2_X}{m^2_Y-m^2_X}\ln\fr{m^2_Y}{m^2_X}-2+ t^2_W\ln\fr{m^2_Y}{m^2_X}\right),\\
 U &=&\fr 1 \pi\left[\left(\fr 3 4 +\frac{m^2_Y}{m^2_Z}+2s^2_W\right)\bar{F}_0(m^2_Z,m_Y,m_Y)+\left(\fr 3 4 +\frac{m^2_X}{m^2_Z}\right)\bar{F}_0(m^2_Z,m_X,m_X)\right.\crn
 &&- \left(\fr 3 2+\frac{m^2_Y+m^2_X}{m^2_W}-\frac{(m^2_Y-m^2_X)^2}{2m^4_W}\right)\bar{F}_0(m^2_W,m_Y,m_X)\crn
 &&-\left(\frac{m^2_Y+m^2_X}{2m^2_W}+\frac{5(m^2_Y-m^2_X)^2}{2m^4_W}+2\right)\bar{F}_0(0,m_Y,m_X)-\frac{5(m^2_Y+m^2_X)}{4m^2_W}\crn
&&-\left.\frac{2(m^2_Y-m^2_X)^2}{m^4_W}-\left(\frac{m^2_Y-m^2_X}{2m^2_W}-\frac{m^4_Y-m^4_X}{m^4_W}-\frac{5(m^4_Y+m^4_X)}{4m^2_W(m^2_Y-m^2_X)}\right)\ln\frac{m^2_Y}{m^2_X} \right],\label{URHN}  \eea
while for the minimal 3-3-1 model, i.e. $\beta=-\sqrt3$, they are  \cite{Sasaki:1992np,Frampton:1997in} 
\bea S &=& \frac{3}{\pi}\left[\fr 7 4 \ln\frac{m^2_Y}{m^2_X}+\left(\fr 3 4+\fr{2(1+2s^2_W)}{3}-\frac{m^2_Y}{m^2_Z}\right)\bar{F}_0(m^2_Z,m_Y,m_Y)\right.\crn
&&-\left.\left(\fr 3 4+\fr{1-4s^2_W}{3}-\frac{m^2_X}{m^2_Z}\right)\bar{F}_0(m^2_Z,m_X,m_X) \right],\label{SMin}\\
T &=& \fr{3\sqrt{2}G_F}{16\pi^2\al}\left(m^2_Y+m^2_X-\fr{2m^2_Y m^2_X}{m^2_Y-m^2_X}\ln\fr{m^2_Y}{m^2_X}\right)\crn
 &&+\fr{1}{4\pi s^2_W}\left(\fr{m^2_Y+m^2_X}{m^2_Y-m^2_X}\ln\fr{m^2_Y}{m^2_X}-2+3 t^2_W\ln\fr{m^2_Y}{m^2_X}\right),\\
 U &=&\fr 1 \pi\left[\left(\fr 3 4 -\frac{2m^2_Y}{m^2_Z}+4s^2_W\right)\bar{F}_0(m^2_Z,m_Y,m_Y)+\left(\fr 3 4 -\frac{2m^2_X}{m^2_Z}-2s^2_W\right)\bar{F}_0(m^2_Z,m_X,m_X)\right.\crn
 &&- \left(\fr 3 2-\frac{2(m^2_Y+m^2_X)}{m^2_W}-\frac{(m^2_Y-m^2_X)^2}{2m^4_W}\right)\bar{F}_0(m^2_W,m_Y,m_X)\crn
 &&+\left(\frac{3(m^2_Y+m^2_X)}{2m^2_W}+\frac{3(m^2_Y-m^2_X)^2}{2m^4_W}-2\right)\bar{F}_0(0,m_Y,m_X)+\frac{3(m^2_Y+m^2_X)}{4m^2_W}\crn
&&-\left.\frac{(m^2_Y-m^2_X)^2}{m^4_W}-\left(\frac{m^2_Y-m^2_X}{4m^2_W}-\frac{m^4_Y-m^4_X}{2m^4_W}+\frac{3m^2_Ym^2_X}{2m^2_W(m^2_Y-m^2_X)}\right)\ln\frac{m^2_Y}{m^2_X} \right]. \label{UMin} \eea  Above, the function $F_0$ is defined as in Ref. \cite{Long:1999bny}. 

\begin{figure}[!h]
\begin{center}
	\includegraphics[scale=0.28]{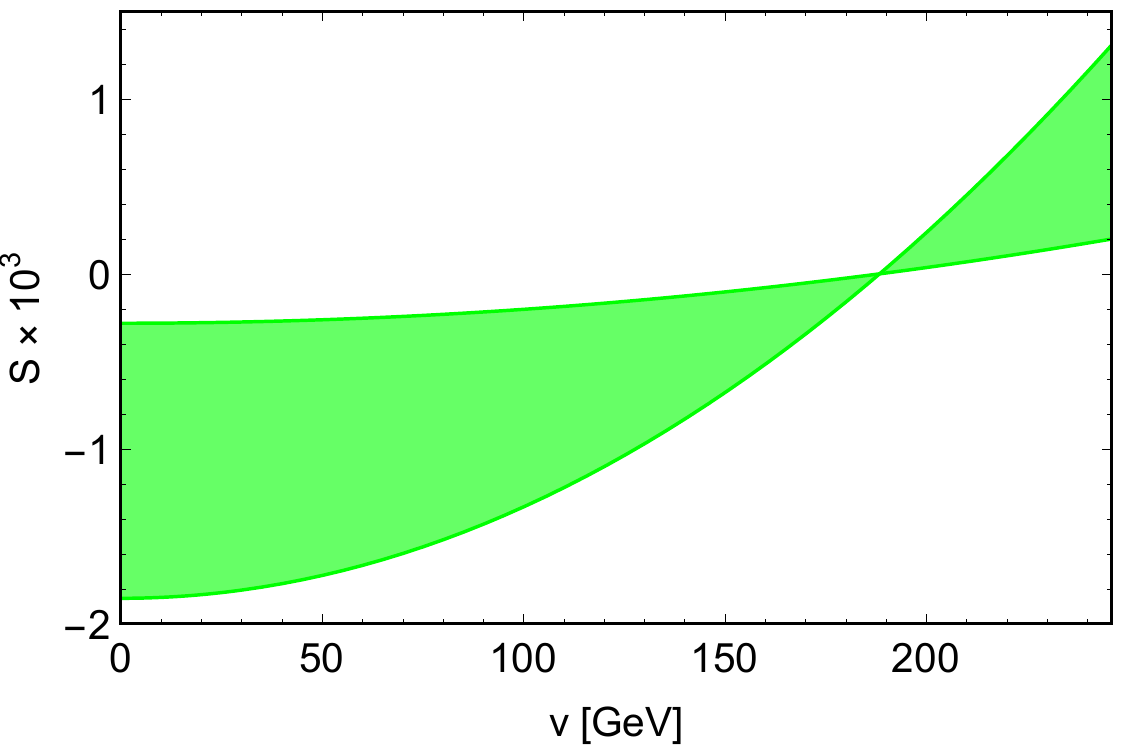}
	\includegraphics[scale=0.28]{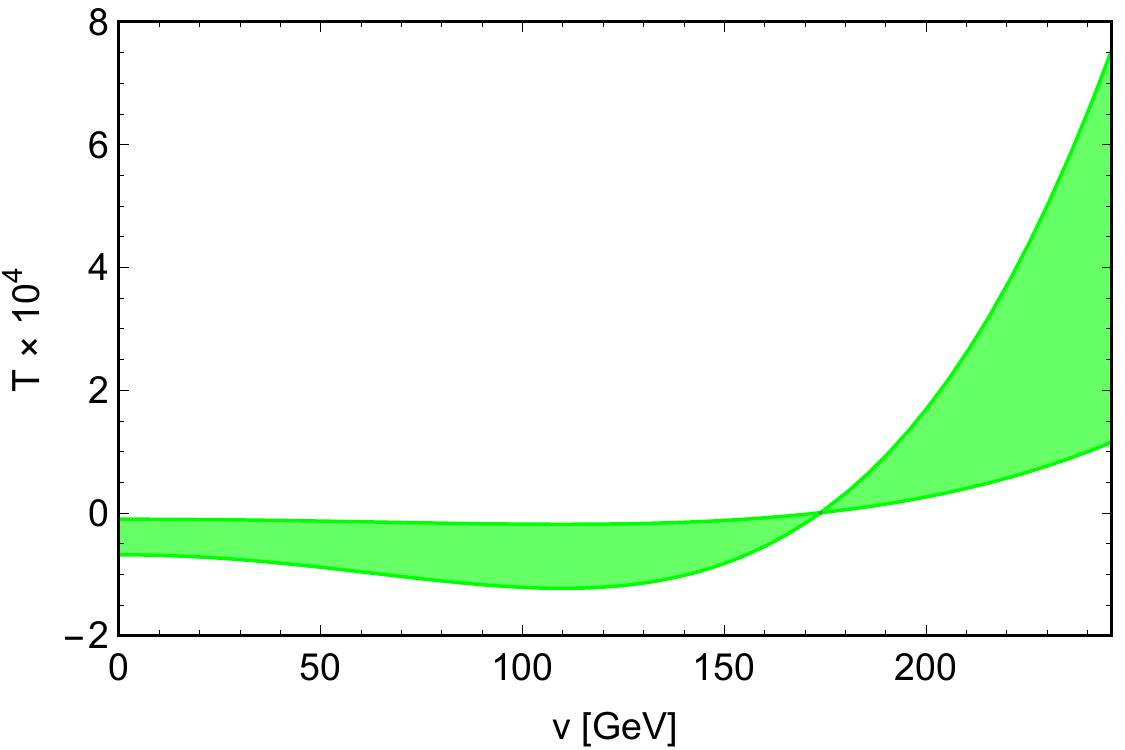}
	\includegraphics[scale=0.28]{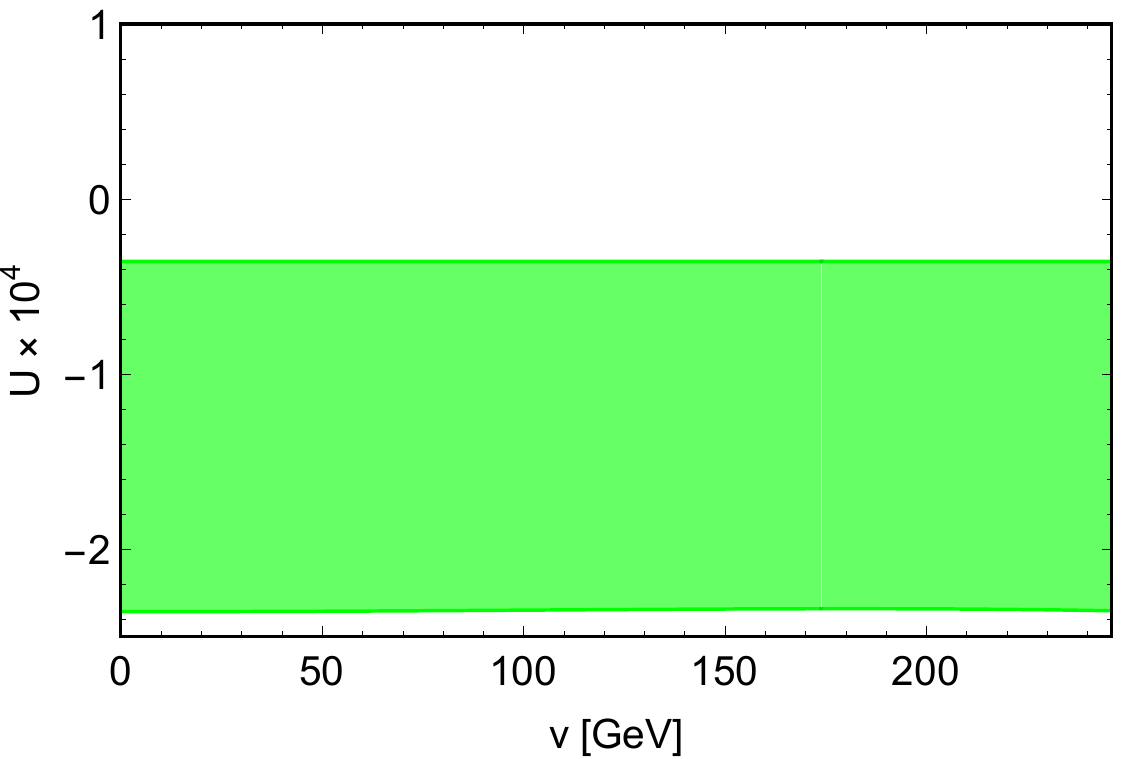}
	\includegraphics[scale=0.28]{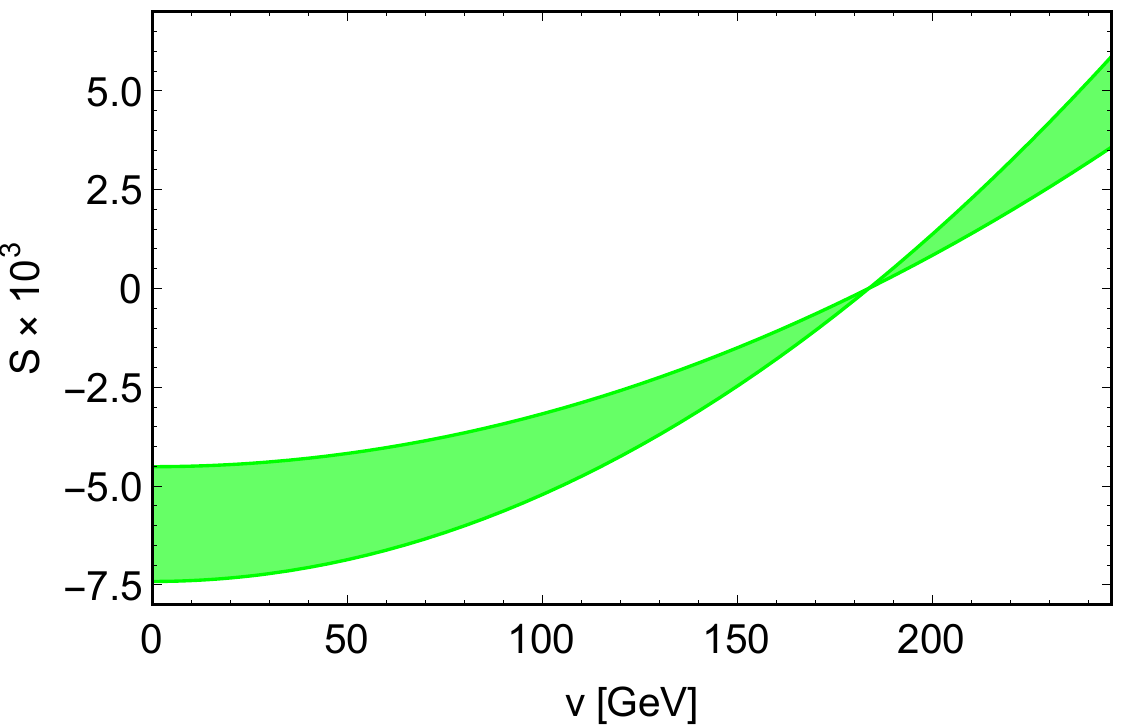}
	\includegraphics[scale=0.28]{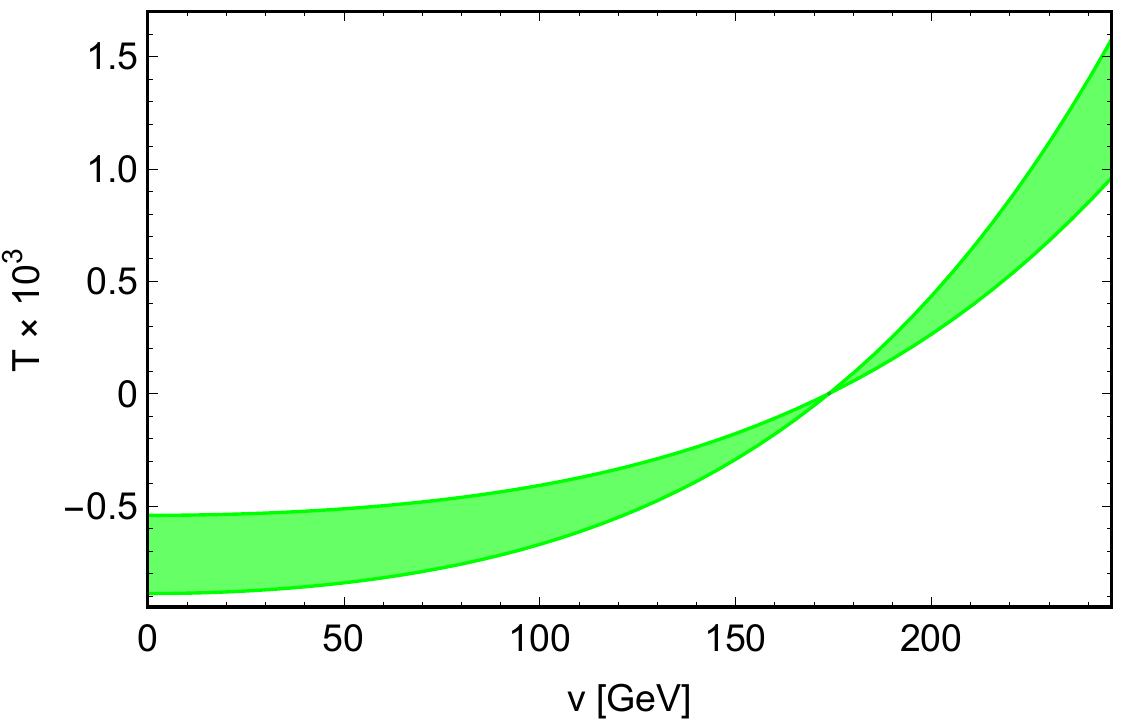}
	\includegraphics[scale=0.28]{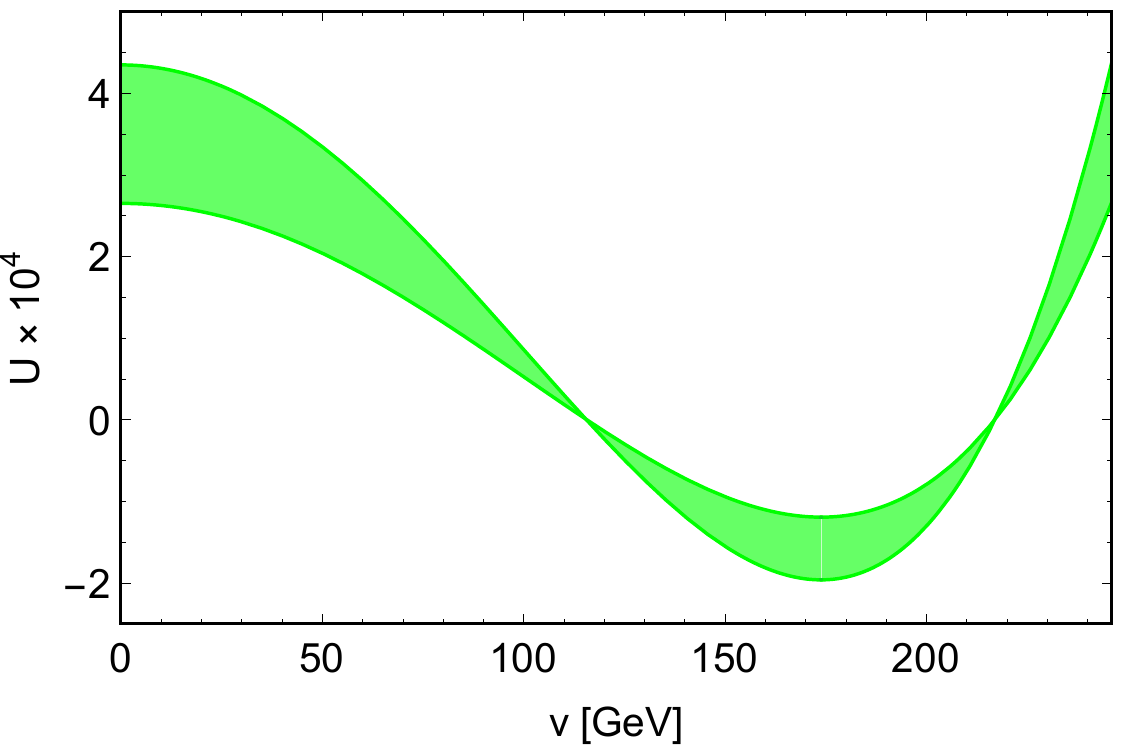}
	\caption{\label{STU} $S$, $T,$ and $U$ parameters plotted as functions of $v$, where the upper panels (the lower panels) correspond to the 3-3-1 model with $\beta=-1/\sqrt3$ ($\beta=-\sqrt3$), and $w$ is free to float in the range $3.9$--10 TeV ($3.9$--5 TeV), as appropriate.}	
\end{center}
\end{figure}

In Fig. \ref{STU}, we plot the $S,\, T,\, U$ parameters as functions of $v$ (where $u$ is followed by $u=\sqrt{(246\text{ GeV})^2-v^2}$), where three upper panels correspond to the 3-3-1 model with $\beta=-1/\sqrt3$ for $w$ in the range of $3.9-10$ TeV, while three lower panels correspond to the 3-3-1 model with $\beta=-\sqrt3$ for $w$ in the range of $3.9-5$ TeV since this model is limited by the Landau pole (cf. Sec. \ref{Constrain}). It is clear that the values of $S,\, T, \, U$ can be negative or positive (except for the rightmost upper panel), depending on the value of $v$ as well as the sign and magnitude of $X$ and $Y$ mass splitting. Additionally, the magnitudes of $S,\, T, \, U$ parameters are very small compared with the value of the $T$ parameter $(T\sim 0.18)$ in the case of tree-level $Z$-$Z'$ and $Z$-$Z'$-$X$ mixings. Hence, the oblique contributions due to the $X,Y$ gauge vector doublet are not significant, i.e. not enough to accommodate the CDF $W$-boson mass anomaly if the tree-level sources of mixing are not included.  

\begin{figure}[!h]
\begin{center}
	\includegraphics[scale=0.4]{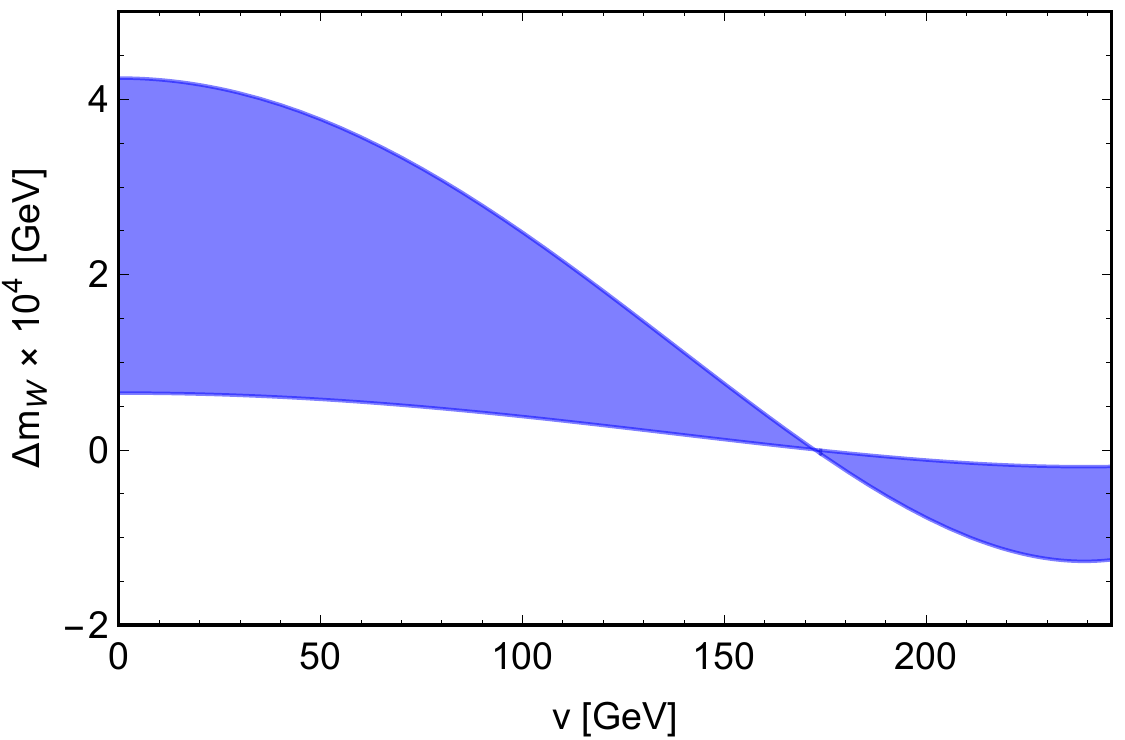}
	\includegraphics[scale=0.4]{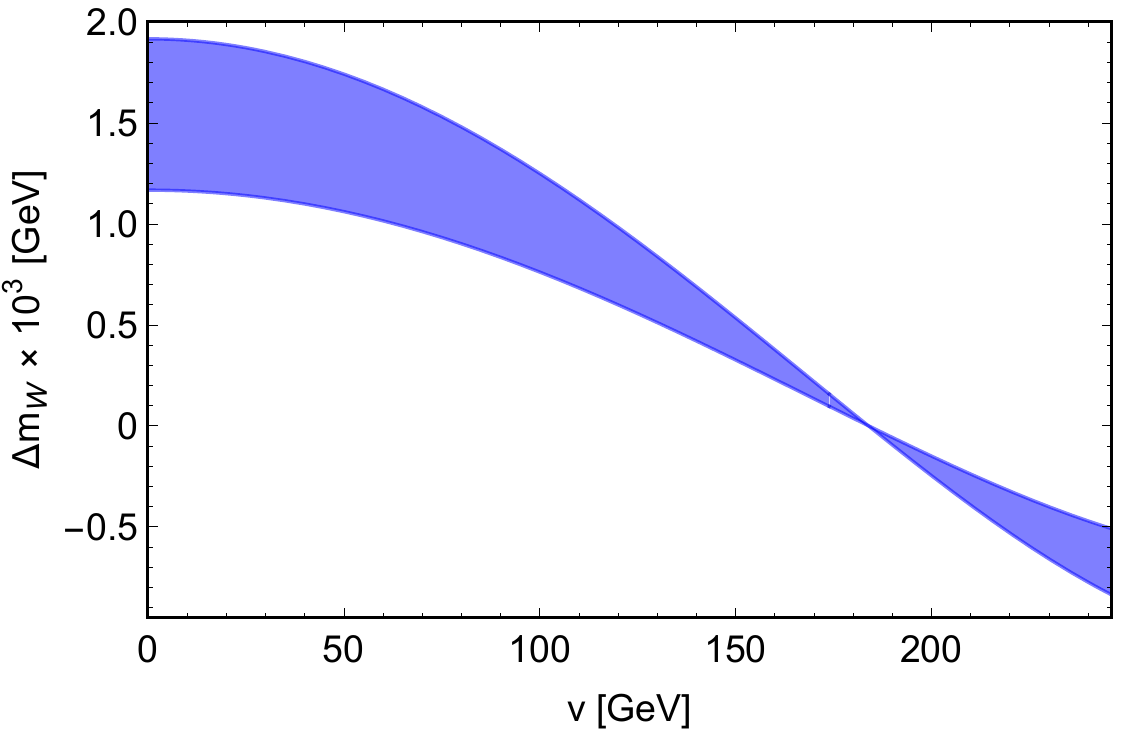}
	\caption{\label{loi1} Contribution of the non-degenerate gauge vector doublet to the $W$ mass shift.}	
\end{center}
\end{figure}

Indeed, substituting (\ref{SRHN})--(\ref{URHN}) [(\ref{SMin})--(\ref{UMin})] into (\ref{WmassSTU}) and taking $v=0$--246 GeV and $w=3.9$--10 TeV ($w=3.9$--5 TeV) for the 3-3-1 model with $\beta=-1/\sqrt3$ ($\beta=-\sqrt3$), we obtain the enhancement of the $W$-boson mass caused by the oblique corrections to be too small, namely, $\Delta m_W\leq 0.00042$ GeV ($\Delta m_W\leq 0.00192$ GeV), as explicitly shown in Fig. \ref{loi1}, incompatible with the experimental measurement.

\subsection{Oblique contributions of a non-degenerate scalar doublet}

Under the standard model symmetry, the three scalar triplets as given correspondingly contain three $SU(2)_L$ scalar doublets, since each triplet is decomposed as $3=2\oplus 1$. One of the doublets is eaten by $W,Z$ leaving only the usual physical Higgs field, other one of the doublets is completely eaten by the $X,Y$ gauge vector doublet. The remaining scalar doublet is really a new physical Higgs doublet, which potentially contributes to the $W$-boson mass via $S,T,U$ parameters, partly noted in \cite{VanDong:2015ifg}. Such a physical Higgs doublet also exits in the 3-3-1 models for dark matter as the first and second entries of inert scalar triplets \cite{Dong:2013ioa,Dong:2014esa}, or in 3-3-1 models with flavor symmetries as contained in scalar flavon triplets \cite{Dong:2010gk,Dong:2010zu,Dong:2011vb}. Even if one includes a scalar sextet \cite{Foot:1992rh,Dong:2008sw,Dong:2010gk,Dong:2010zu,Dong:2011vb}, a new Higgs doublet correspondingly arises, since $6=3\oplus 2\oplus 1$ under $SU(2)_L$. In this case, the $SU(2)_L$ scalar triplet also contributes to $S,T,U$ parameters, but this contribution is neglected because of its significant tree-level contribution as discussed below. 

That said, a new physical Higgs doublet in addition to the usual Higgs doublet is popularly presented in the 3-3-1 model. Without loss of generality, we consider a generic scalar triplet,
\be \phi = (\phi_1,\, \phi_2,\, \phi_3)^T\sim (1,3,X_\phi),\ee
which contains the new $SU(2)_L$ Higgs doublet, i.e. $(\phi_1,\phi_2)$, as desirable. Here $\phi$ can have an arbitrary $X$-charge, making a significant contribution to the oblique parameters independent of electric charge. It is easily shown that the contributions of $\phi$ to $S,U$ parameters are more smaller than that to $T$, if the scalar fields are radically heavier than $W,Z$ masses. Indeed, in the 3-3-1 model, both $\phi_{1,2}$ are typically heavy at TeV scale, while the mass-squared splitting of $\phi_{1,2}$ is only proportional to the weak scale \cite{Dong:2013ioa,Dong:2014esa}. In this case, the $W$-boson mass shift is governed by the $T$ parameter, such as
\be \Delta m^2_W = \frac{c_W^4 m_Z^2}{c_W^2-s_W^2} \fr{G_F}{8\sqrt{2}\pi^2}\left(m^2_1+m^2_2-\fr{2m^2_1 m^2_2}{m^2_1-m^2_2}\ln\fr{m^2_1}{m^2_2}\right), \ee where $m_{1,2}$ are the masses of $\phi_{1,2}$, respectively.

\begin{figure}[!h]
\begin{center}
	\includegraphics[scale=0.45]{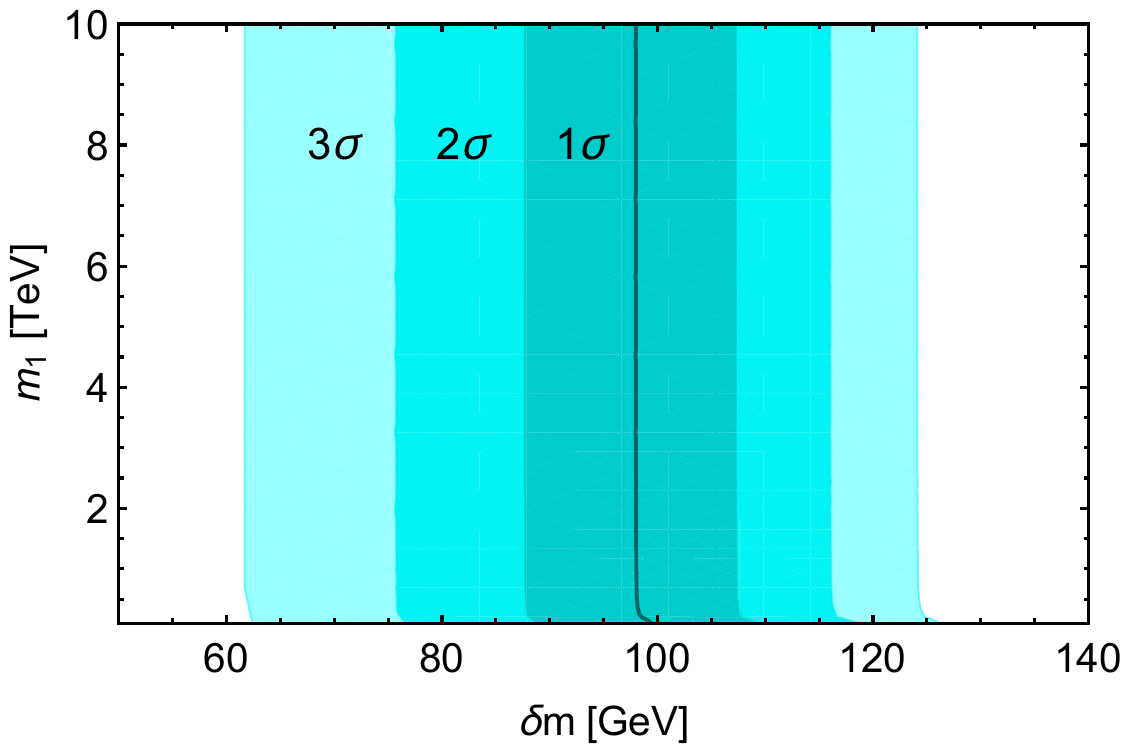}
	\caption{\label{Inert} CDF-allowed regions contoured in $\delta m$--$m_1$ plane.}	
\end{center}
\end{figure}

Taking the central values of $W$ mass from the CDF experiment \cite{CDF:2022hxs} and the standard model prediction \cite{ParticleDataGroup:2020ssz}, respectively, in Fig. \ref{Inert} we contour $\Delta m^2_W$ as function of $\delta m=m_2-m_1$ and $m_1$ by the black line. We obtain the viable value, $\delta m \simeq 98$ GeV, as appropriate. Further, the contributions to $\Delta m^2_W$ depend only on the mass splitting $\delta m$, nearly insensitive to $m_1$. For clarity, the $1\sigma$, $2\sigma$, and $3\sigma$ ranges of the measured $W$ mass are also shown (in cyans).

Last, but not least, it is clear that $m_1\sim \la_{\phi-H} v^2/\delta m$ where $\la_{\phi-H}$ relates to the coupling of $\phi$ to the usual Higgs field. Hence, $\phi$ obtains a mass at TeV, i.e. $m_1\sim $ TeV, only if $\la_{\phi-H}$ is at the perturbative limit \cite{VanDong:2022rox}.

\section{\label{sextet} Tree-level contribution of a sextet}

A scalar sextet of type,
\be S=\begin{pmatrix} 
S^0_{11} & \fr{S^-_{12}}{\sqrt{2}} & \fr{S^q_{13}}{\sqrt{2}}\\
\fr{S^-_{12}}{\sqrt{2}} & S^{--}_{22} & \fr{S^{q-1}_{23}}{\sqrt{2}}\\
\fr{S^q_{13}}{\sqrt{2}} & \fr{S^{q-1}_{23}}{\sqrt{2}} & S^{2q}_{33}
\end{pmatrix},\ee often studied in the 3-3-1 models \cite{Valle:1983dk,Foot:1992rh,Tully:2000kk,Ky:2005yq,Dong:2008sw}, can develop a vacuum value such as
\be \langle S\rangle =\fr{1}{\sqrt{2}}\begin{pmatrix} 
\kappa & 0 & 0\\
0 & 0& 0\\
0 & 0 & 0
\end{pmatrix}.\ee Other fields if electrically neutral can also have a VEV. But they belong to a $SU(2)_L$ doublet or singlet, giving a contribution similar to the above cases of scalar triplets, and are not interested. Here, the nontrivial vacuum structure is associated with the $SU(2)_L$ scalar triplet, $(S_{11}, S_{12}, S_{22})$, contained in the sextet, unlike those in the cases of scalar triplets. Unfortunately, this sextet gives a negative contribution to the $\rho$-parameter, 
\be \rho\simeq 1-\fr{2\kappa^2}{u^2+v^2}, \ee as shown in \cite{Dong:2008sw}, incompatible with the experiment. This case is similar to a scalar triplet in the type II seesaw mechanism added to the standard model.

However, if we consider an alternative scalar sextet of type, 
\be \sigma=\begin{pmatrix} 
\sigma^+_{11} & \fr{\sigma^0_{12}}{\sqrt{2}} & \fr{\sigma^{q+1}_{13}}{\sqrt{2}}\\
\fr{\sigma^0_{12}}{\sqrt{2}} & \sigma^{-}_{22} & \fr{\sigma^{q}_{23}}{\sqrt{2}}\\
\fr{\sigma^{q+1}_{13}}{\sqrt{2}} & \fr{\sigma^{q}_{23}}{\sqrt{2}} & \sigma^{2q+1}_{33}
\end{pmatrix},\ee studied in the 3-3-1 model for dark matter \cite{Dong:2014esa}, the $SU(2)_L$ scalar triplet $(\sigma_{11},\sigma_{12},\sigma_{22})$ contained in the sextet develops a VEV in different way, such as 
\be \langle \sigma\rangle =\fr{1}{2}\begin{pmatrix} 
0 & \kappa' & 0\\
\kappa' & 0& 0\\
0 & 0 & 0
\end{pmatrix}.\ee It gives a positive contribution to the $\rho$-parameter, such as 
\be \rho\simeq 1+\fr{4\kappa'^2}{u^2+v^2}. \ee Comparing to the $W$-mass shift, $\kappa'$ is about 4.5 GeV, similarly in size to $u'$ bounded above. This case is similar to a scalar triplet with $Y=0$ added to the standard model \cite{Cheng:2022jyi}.  

It is stressed that the mentioned scalar sextets contribute to the gauge boson mass spectra differently from the scalar triplets in previous sections. However, the results obtained according to the scalar triplets are easily generalized for the sextets with the aid of \cite{Dong:2005wgt}.

\section{\label{Constrain}Existing bounds}

We now present the running coupling and Landau pole limit which place a constraint on $\beta,q$ parameters. We also discuss FCNCs which reveal important information on $B-L$-violating VEVs and new physics scale. Last, but not least, we examine collider bounds, validating the previous constraints, as well as we summarize the relevant bounds used for model classification and updated with $W$-mass measurement.        

\subsection{\label{beqlpdt}Running coupling and Landau pole}

A gauge coupling commonly denoted as $g=\{g_s,g,g_X\}$ changes with renormalization scale $\mu$ through the RG equation, \be \fr{\pa g}{\pa \ln \mu}=\beta(g)=-\fr{g^3}{16\pi^2}b_g,\ee where the 1-loop beta function is given by \be b_g=\fr{11}{3} C_V-\fr 2 3 \sum_L C_L-\fr 2 3 \sum_R C_R -\fr 1 3 \sum_S C_S,\ee where $V,L/R,S$ indicate vector, left/right fermion, and scalar field representations under the relevant gauge group, respectively. For $SU(3)$, $C_V=3$ and $C_L=C_R=C_S=1/2$ for (anti)triplets, while for $U(1)_{X}$, $C_V=0$ and $C_{L,R,S}=X^2_{L,R,S}$ for $L, R, S$ fields, respectively. We obtain $b_{g_s}=5>0$ and $b_g=13/2>0$. Thus, $g_s,g$ decrease when $\mu$ increases. There is no Landau pole associated with $g_s,g$. However, because of $C_V=0$ and $C_{L,R,S}=X^2_{L,R,S}>0$, we always have $b_{g_X}<0$. Hence, $g_X$ increases when $\mu$ increases. 

In contrast to the standard model and grand unification, a finite Landau pole associated with $g_X$ potentially arises because of $U(1)_X$ along with $SU(3)_L$ embedded in $U(1)_Q$. Indeed, as given before, the electric charge operator defines the photon field that couples to it and that the normalization of the photon field implies \be s^2_W=\fr{g^2_X}{g^2+g^2_X(1+\beta^2)}<\fr{1}{1+\beta^2},\ee where note that $g$ is always finitely nonzero. When the energy scale increases, $g_X/g$ increases as long as $s^2_W$ approaches the r.h.s of the inequality. The model encounters a Landau pole $\mu$ at which $s^2_W(\mu)=\fr{1}{1+\beta^2}$, or equivalently $g_X(\mu)= \infty$, where the scale of $\mu$ depends heavily on $\beta$ value. It is stressed that the model is valid only if the Landau pole $\mu$ is larger than the new physics scale, i.e. $\mu> w$, thus than the weak scale $u,v$. Correspondingly, we have $s^2_W(\mu)=\fr{1}{1+\beta^2}>s^2_W(u,v)$. This yields $|\beta|<\cot_W(u,v)\simeq 1.824$ for $s^2_W(u,v)\simeq 0.231$, which translates to $-2.08 < q < 1.08$, due to $\beta=-(1+2q)/\sqrt{3}$. 

The electric charge of $E$ is very constrained, taking $q=-2,-1,0,1$ for integer charges. For $q=-2,1$, we have $|\beta|=\sqrt{3}$ and $s^2_W(\mu)=1/4$ just above that at the weak scale. In this case, the model presents a low Landau pole at $\mu=4$--5 TeV, as shown in \cite{Dias:2004dc,Dias:2004wk} for the minimal 3-3-1 model. However, if $q=-1,0$, which include the 3-3-1 model with heavy charged leptons and the 3-3-1 model with right-handed neutrinos, we have $|\beta|=1/\sqrt{3}$ and $s^2_W(\mu)=3/4$ that is much beyond the typical value $s^2_W=3/8$ at the grand unification scale, close to the Planck regime. In this case, the model still possesses a Landau pole, but this pole is beyond the Planck scale. 

\subsection{\label{fcncsdt}FCNCs}

There are two sources for FCNCs in the 3-3-1 model that come from the nonuniversality of quark families under the gauge group and the possible mixing of ordinary quarks and exotic quarks, respectively. The former occurs for every $q$ associated with $Z'$ current, while the latter arises only if $q=0$ or $-1$ concerning $Z$ current. Particularly for $q=0$, it happens with the 3-3-1 model with right-handed neutrinos but not with the 3-3-1 model with neutral (heavy) fermions. These FCNCs are actually induced at tree-level and indeed dangerous. The model with $q=-1$ happening similarly to the case $q=0$, as well as the FCNCs that are possibly arisen/coupled to the Higgs and new Higgs fields, will not be discussed.             

In the 3-3-1 model with $q=0$, the ordinary and exotic quarks of up-type $(u_a,J_3)$ and of down-type $(d_a,J_\al)$ each mix by themselves due to the lepton-number violating VEVs of $\eta_3,\chi_1$ as well as the lepton-number violating Yukawa couplings (called $s$'s) that like ordinary Yukawa couplings (called $h$'s) but appropriately interchange the right-handed components of ordinary and exotic quarks, as supplied above \cite{Dong:2008sw}. We define mixing matrices, $(u_1\ u_2\ u_3\ J_3)^T_{L,R}=V_{uL,R}(u\ c\ t\ T)^T_{L,R}$
and $(d_1\ d_2\ d_3\ J_1\ J_2)^T_{L,R}=V_{dL,R}(d\ s\ b\ B_1\ B_2)^T_{L,R}$, such that the $4\times 4$ mass matrix of $(u_a,J_3)$ and the $5 \times 5$ mass matrix of $(d_a,J_\al)$ are diagonalized. Because the ordinary and exotic quarks have different $T_3$ weak isospin, the tree-level FCNCs of $Z$ boson arise, such as \be \mathcal{L}\supset (\pm)\fr{g}{2c_W}\bar{q}_{iL}\ga^\mu q_{jL} (V^*_{qL})_{Ii}(V_{qL})_{Ij}Z_\mu,\ee i.e. the standard model CKM mechanism does not work \cite{Dong:2014wsa}. Here, we denote $q$ as either $u$ for up-type or $d$ for down-type quarks, which should {\it not} be confused with the $q$ charge parameter used throughout, $i,j=1,2,3$ label ordinary physical quarks, and $I=J_3$ and plus sign applies for $V_{u}$, while $I=J_\al$ and minus sign applies for $V_d$. Integrating $Z$ out as well as using $m^2_Z\simeq (g^2/4c^2_W)(u^2+v^2)$, we obtain effective interactions, \bea \mathcal{H}^{\mathrm{NP}}_{\mathrm{eff}} &\supset& (\bar{q}_{iL}\ga^\mu q_{jL})^2[(V^*_{qL})_{Ii}(V_{qL})_{Ij}]^2\fr{1}{u^2+v^2}\crn
&\supset& \Delta C_{\bar{d}s}(\bar{d}_L\ga^\mu s_L)^2+\cdots,\eea where $\Delta C_{\bar{d}s}\equiv [(V^*_{dL})_{I1}(V_{dL})_{I2}]^2/(u^2+v^2)$, and ``$\cdots$'' indicates its conjugate as well as other four-quark systems. The strongest bound comes from the $K^0$--$\bar{K}^0$ mixing that requires the relevant coupling, $\Delta C_{\bar{d}s}$, to be smaller than $(10^4\ \mathrm{TeV})^{-2}$ \cite{ParticleDataGroup:2020ssz}, implying  \be |(V^*_{dL})_{I1}(V_{dL})_{I2}|\lesssim 10^{-5}.\ee The mixing of ordinary and exotic quarks is much smaller than the smallest mixing element of CKM matrix (around $5\times 10^{-3}$), which may be understood in the 3-3-1-1 model \cite{Dong:2013wca}. For the present model, we safely assume $(V_{qL})_{Ii}\sim u'/ u \sim w' / w \sim s/h\sim 10^{-2}$--$10^{-3}$.\footnote{A basis changing so that either $\langle \eta_3\rangle =0$ or $\langle \chi_1\rangle =0$ as studied in the literature also changes quark states, as a result. Hence, this condition is generically applied.} 

It is noted that the model with $q\neq 0,-1$ has only FCNCs associated with nonuniversal $Z'$ couplings, because the ordinary and exotic quarks do not mix. The following computation would apply for all cases of $q$-charge, since for $q=0$ (or $-1$) the ordinary and exotic quark mixing negligibly contributes to this kind of the FCNCs, as shown in \cite{Dong:2014wsa}. Because the third family of quarks transforms under $SU(3)_L\otimes U(1)_X$ differently from the first two, there must be tree-level FCNCs. Indeed, using $X=Q-T_3-\beta T_8$, the interaction of the neutral current is $\mathcal{L}\supset -g\bar{F}\ga^\mu[T_3 A_{3\mu} + T_8 A_{8\mu}+t_X(Q-T_3-\beta T_8)B_\mu]F$, where $F$ runs over fermion multiplets. It is clear that the terms of $T_3$ and $Q$, as well as all terms of $\nu_a$, $e_a$, $E_a$, $J_\al$, and $J_3$, do not flavor change. The relevant part includes only $T_8$ with ordinary quarks, such as $\mathcal{L}\supset -g\bar{q}_L\ga^\mu T_{8q}q_L (A_{8\mu}-\beta t_X B_\mu)
=-g\bar{q}_L\ga^\mu T_{8q} q_L Z'_\mu/\sqrt{1-\beta^2 t^2_W}$,
where $q$ denotes either up-type or down-type quarks and $T_{8q}=\fr{1}{2\sqrt{3}}\mathrm{diag}(-1,-1,1)$ combines their $T_8$ charge. Changing to the mass basis, $q_{aL,R}=(V_{qL,R})_{ai}q_{iL,R}$, we have  
\bea \mathcal{L} \supset -\fr{g}{\sqrt{3}\sqrt{1-\beta^2 t^2_W}}\bar{q}_{iL}\ga^\mu q_{jL} (V^*_{qL})_{3i}(V_{qL})_{3j}Z'_\mu, \eea which implies FCNCs for $i\neq j$. Integrating $Z'$ out and using $m^2_{Z'}\simeq g^2 w^2/3(1-\beta^2 t^2_W)$, we obtain effective interactions, 
\bea \mathcal{H}^{\mathrm{NP}}_{\mathrm{eff}} &\supset& (\bar{q}_{iL}\ga^\mu q_{jL})^2 [(V^*_{qL})_{3i}(V_{qL})_{3j}]^2\fr{1}{w^2}\crn
&\supset& \Delta C'_{\bar{s}b}(\bar{s}_L\ga^\mu b_L)^2+\cdots,\eea where $\Delta C'_{\bar{s}b}\equiv [(V^*_{dL})_{32}(V_{dL})_{33}]^2/w^2$, and ``$\cdots$'' stands for its conjugate and other four-quark systems. The strongest bound comes from the $B^0_s-\bar{B}^0_s$ system, given by \cite{ParticleDataGroup:2020ssz} 
\be [(V^*_{dL})_{32}(V_{dL})_{33}]^2\fr{1}{w^2}<\fr{1}{(100\ \mathrm{TeV})^2}.\ee Assuming $V_{uL}=1$, the CKM factor is $|(V^*_{dL})_{32}(V_{dL})_{33}|\simeq 3.9\times 10^{-2}$, which translates to $w>3.9$ TeV. This bound is independent of $\beta$, i.e. applying for every 3-3-1 model. Last, notice that the $K^0$--$\bar{K}^0$ mixing gives a bound, $w>3.6$ TeV, slightly smaller than the given one, which has not been signified.     

\subsection{\label{colisdt0}Collider searches}

The LEPII studied the process $e^+ e^-\rightarrow f\bar{f}$, where $f$ is an ordinary fermion, through exchange of a new heavy gauge boson $Z'$, described by effective interactions,      
\be \mathcal{L}_{\mathrm{eff}}\supset \fr{g^2}{c^2_Wm^2_{Z'}}\left[\bar{e}\ga^\mu(a^{Z'}_L(e)P_L+a^{Z'}_R(e)P_R)e\right]\left[\bar{f}\ga_\mu(a^{Z'}_L(f)P_L+a^{Z'}_R(f)P_R)f\right],\ee where $a^{Z'}_{L,R}(f)=\fr 1 2 [g^{Z'}_V(f)\pm g^{Z'}_A(f)]$, as usual. Considering $f=\mu,\tau$, the charged leptons have equal couplings, and we rewrite 
\be \mathcal{L}_{\mathrm{eff}}\supset \fr{g^2[a^{Z'}_L(e)]^2}{c^2_W m^2_{Z'}}(\bar{e}\ga^\mu P_L e)(\bar{f}\ga_\mu P_Lf)+(LR)+(RL)+(RR),\ee where the successive terms differ from the first one only in chiral structure, and 
\be a^{Z'}_L (e) = \fr{c_W+\sqrt{3}\beta s_W t_W}{2\sqrt{3}\sqrt{1-\beta^2 t^2_W}},\hs a^{Z'}_R(e)=\fr{\beta s_W t_W}{\sqrt{1-\beta^2 t^2_W}}. \ee 

The LEPII supplied bounds for such chiral couplings, typically \cite{ALEPH:2006bhb}  
\be \fr{g^2[a^{Z'}_L(e)]^2}{c^2_W m^2_{Z'}} <\fr{1}{(6\ \mathrm{TeV})^2}.\ee Using $m^2_{Z'}\simeq g^2w^2/[3(1-\beta^2t^2_W)]$, we get 
\be w> 3\times (1+\sqrt{3}\beta t^2_W)\ \mathrm{TeV},\ee which are 5.7 TeV, 3.9 TeV, 2.1 TeV, and 0.3 TeV, for $q=-2$, $-1$, 0, and 1, respectively. The first case is ruled out by the Landau pole, while the last case means that $Z'$ negligibly contributes to the process, since $w$ would be at TeV due to the FCNCs above. That said, the 3-3-1 model with $q=0$ and $\pm 1$ would be viable, as taken into account.  

The LHC searched for dilepton signals through the process $pp\to ff^c$ that is contributed by $Z'$, supplying a bound $m_{Z'}\sim 4$ TeV for $Z'$ couplings identical to those of the usual $Z$ boson \cite{ATLAS:2017fih}. It is noticed that $Z'$ in our model couples similarly to $Z$, governed by the common $g$ coupling but having a small difference due to $\beta$. Therefore, the bound as given applies to our model with some extent since the couplings are not identical. The $Z'$ mass limit thus converts to a $w$ bound comparable to the LEPII.                    

\subsection{Summary of the existing bounds with updates for model building and $W$ mass}

The viable range of $\beta,q$ and the Landau pole limit determined in Sec. \ref{beqlpdt} have appropriately been supplied to Table \ref{331versions} for classifying 3-3-1 versions. Additionally, all the relevant, existing bounds of this section, i.e. Secs. \ref{beqlpdt}, \ref{fcncsdt}, and \ref{colisdt0}, have appropriately been combined with the $W$-mass measurement in the previous sections. 

For convenience in reading, we give a summary of all the bounds obtained in this section as well as their application to the $W$-mass deviation in the previous sections, as in Table~\ref{tabdt}. That said, the constraints previously given concerning the $W$ mass, such as $(w,v)$ that results from Figs. \ref{CDF1}, \ref{STU}, and \ref{loi1}, are indeed the updated bounds, while the conditions $u'\ll u$ and $w'\ll w$ which are obviously used for the gauge spectrum and the $W$-mass shift are suitable to the bound in this section.   

\begin{table}[h]
\bc
\begin{tabular}{|l|l|l|l|}
\hline\hline
Parameter & Existing bounds & Special values (respectively) & Updated with $W$-mass \\
\hline
$q$ & $(-2.08,1.08)$ & $-2$,\hs $-1$,\hs 0,\hs 1 & $\beta=\sqrt{3}$ excluded, since $w>\mu$;\\ \cline{1-3} 
$\beta=-\fr{1+2q}{\sqrt{3}}$ & $(-1.824,1.824)$ & $\sqrt{3}$,\hs $\fr{1}{\sqrt{3}}$,\hs $-\fr{1}{\sqrt{3}}$,\hs $-\sqrt{3}$ & $\beta=\fr{1}{\sqrt{3}},-\fr{1}{\sqrt{3}},-\sqrt{3}$ interested \\ \hline
Landau pole & Weak/TeV to much & \multirow{2}{*}{4--5 TeV, $>M_{\mathrm{Pl}}$, $>M_{\mathrm{Pl}}$, 4--5 TeV} & $\mu=5$ TeV suitably applied to \\ 
($\mu$) & beyond Planck scale & & Figs. \ref{CDF1}, \ref{STU}, and \ref{loi1} \\ \hline
Normal VEVs & \multicolumn{2}{c|}{$u^2+v^2=(246\ \mathrm{GeV})^2$,\hs $\mu>w> 3.9$ TeV for $\beta\leq \fr{1}{\sqrt{3}}$} & $w=3.9$ TeV bounds actually  \\ 
$(u,v,w)$ & \multicolumn{2}{c|}{or $\mu>w>3(1+\sqrt{3}\beta t^2_W)> 3.9$ TeV for $\beta > \fr{1}{\sqrt{3}}$} & included to Figs. \ref{CDF1}, \ref{STU}, and \ref{loi1} \\ \hline
Violating param. &\multicolumn{2}{c|}{\multirow{2}{*}{$\fr{u'}{u}\sim \fr{w'}{w}\sim \fr{s}{h}\sim 10^{-2}$--$10^{-3}$}} & Used appropriately to Sec. \ref{iicdt} \\ 
$(u',w',s)$ & \multicolumn{2}{c|}{} & as well as Eqs. (\ref{W1mass}) and (\ref{Z1mass}) \\
\hline\hline
\end{tabular}
\caption[]{\label{tabdt} A summary of existing bounds and their updates with $W$-mass result.}
\ec
\end{table}     

\section{\label{conclusion}Conclusion}

We have examined a variety of variants of the 3-3-1 model, characterized by a basic charge parameter $q$ and the behavior of $B-L$ symmetry through its residual matter parity~$P_M$. The 3-3-1 model generally has a standard vacuum structure, except for the 3-3-1 variant with right-handed neutrinos or with heavy charged leptons which can develop an abnormal vacuum structure, not protected by $P_M$. We have diagonalized the gauge spectra according to the two kinds of vacuum in detail.

There are two kinds of constraints for this model, which are jointly considered and consistently updated:  

\noindent {\it New experiment}: We have investigated the various contributions of the 3-3-1 model to the CDF $W$-mass anomaly, and we conclude that the $Z$--$Z'$ mixing due to the 3-3-1 breakdown by the normal VEVs $u,v,w$ and a non-degenerate physical Higgs doublet popularly existed in the model can explain this anomaly, separately. Additionally, a scalar sextet that contains a $SU(2)_L$ Higgs triplet with $Y=0$ can also solve this puzzle. Furthermore, considering the special 3-3-1 versions with $q=0,\pm 1$, the viable regimes of $u,v,w$ have been derived, obeying the FCNC, collider, and relevant Landau-pole limits. For the case of the heavy non-degenerate Higgs doublet, its mass splitting should be at 98 GeV, whereas for the case of the scalar sextet, the tiny VEV should be about 4.5 GeV.    

\noindent {\it Existing bounds}: The running coupling and Landau pole limit imply a bound for $q$, i.e. $-2.08<q<1.08$. The minimal 3-3-1 model $(q=1)$ and the 3-3-1 model with $q=-2$ have a Landau pole at TeV scale, while the 3-3-1 model with right-handed neutrinos/neutral fermions $(q=0)$ and the 3-3-1 model with heavy charged leptons $(q=-1)$ possess a Landau pole beyond the Planck scale. The FCNCs set a constraint on $B-L$ violating parameters, such as $u'/u\sim w'/w\sim s/h\sim 10^{-2}$--$10^{-3}$, for the 3-3-1 model with right-handed neutrinos, while they imply a bound for the new physics scale $w>3.9$ TeV, valid for every 3-3-1 version. The LEPII gives a strong bound for the 3-3-1 model with $q=-2$, i.e. $w>5.7$ TeV, which is bigger than its Landau pole, $\mu\sim 4$--5 TeV, hence this version is ruled out. The collider bounds for 3-3-1 versions with $q=0,\pm1$ are appropriate to the FCNCs.                 

\section*{Acknowledgments}

This research is funded by Vietnam National Foundation for Science and Technology Development (NAFOSTED) under grant number 103.01-2019.353.

\bibliographystyle{JHEP}
\bibliography{combine}

\end{document}